\pdfoutput=1
\documentclass[letterpaper,twocolumn,10pt]{article}

\usepackage[twoside=true, head=13pt,
     paperwidth=8.5in, paperheight=11in,
     includeheadfoot=false, columnsep=2pc,
     top=1in, bottom=1in, inner=0.75in, outer=0.75in,
     marginparwidth=2pc,heightrounded]{geometry}
\usepackage{cuted}  %
\usepackage{graphicx}
\usepackage[tt=false, type1=true]{libertine}
\usepackage{mathptmx}
\usepackage{amssymb}
\usepackage[varqu]{zi4}
\usepackage[T1]{fontenc}
\usepackage{enumitem}
\usepackage[font={footnotesize},labelfont={footnotesize,bf},textfont={footnotesize,it}]{caption}
\usepackage{hyperref}
\usepackage{xspace}
\usepackage{xcolor}
\usepackage{tikz}
\usetikzlibrary{arrows.meta}
\usepackage{algorithm}
\usepackage{algpseudocode}
\usepackage{amsmath}
\usepackage{caption}
\usepackage{float}

\usepackage{breakurl}
\usepackage[most]{tcolorbox}
\usepackage{booktabs}
\usepackage{multirow}

\newcommand{\proposed}{AISTRAL\xspace}
\newcommand{\ainet}{\emph{AI-Native Internet}\xspace}

\newcommand{\para}[1]{\vspace{1mm}\noindent\textbf{#1}}

\newcommand{\parai}[1]{\vspace{1mm}\textit{#1}}

\newenvironment{smartlist}{%
    \begin{list}{\textbf{$\bullet$}}{%
        \setlength{\topsep}{-\parskip}\addtolength{\topsep}{0pt plus 0.5pt}
        \setlength{\partopsep}{0pt plus 0.5pt}%
        \setlength{\itemsep}{0pt plus 0.5pt}%
        \setlength{\parsep}{0pt plus 0.5pt}%
        \setlength{\leftmargin}{1.3em}%
        \setlength{\labelwidth}{0.7em}%
        \setlength{\labelsep}{0.5em}%
        \setlength{\listparindent}{0pt}%
    }%
}{%
    \end{list}%
}

\newif\ifsubmission
\submissiontrue

\ifsubmission

\newcommand{\zafar}[1]{}
\newcommand{\mbnote}[1]{}
\newcommand{\mcnote}[1]{}
\newcommand{\yxnote}[1]{}
\newcommand{\sfnote}[1]{}
\else

\newcommand{\zafar}[1]{{\footnotesize\color{purple}[ZQ: #1]}}
\newcommand{\mbnote}[1]{{\footnotesize\color{red}[MB: #1]}}
\newcommand{\mcnote}[1]{{\footnotesize\color{orange}[MC: #1]}}
\newcommand{\yxnote}[1]{{\footnotesize\color{blue}[YX: #1]}}
\newcommand{\sfnote}[1]{{\footnotesize\color{olive}[SF: #1]}}
\fi

\usepackage{circledsteps}

\definecolor{aisOrange}{RGB}{190,105,40}
\definecolor{aisBlue}{RGB}{55,95,145}
\definecolor{aisGreen}{RGB}{50,120,100}
\definecolor{aisGray}{RGB}{90,90,90}

\newtcolorbox{insight}[1][]{
  colback=blue!5!white,      %
  colframe=blue!60!black,    %
  fonttitle=\bfseries,
  title={Insight},
  coltitle=white,
  boxrule=0.8pt,
  arc=2pt,
  left=6pt,right=6pt,top=6pt,bottom=6pt,
  #1
}

\title{From Crawled to Published: A Publisher-Controlled Web Architecture for an AI-Native Internet}
\author{%
Yixi Chen\textsuperscript{1}\quad
Muhammad Bilal\textsuperscript{1}\quad
Tie Ma\textsuperscript{2,1}\\
Zafar Ayyub Qazi\textsuperscript{3}\quad
Marco Canini\textsuperscript{1}\quad
Suhaib Fahmy\textsuperscript{1}\\[6pt]
{\normalsize\textsuperscript{1}KAUST\qquad
\textsuperscript{2}Beihang University\qquad
\textsuperscript{3}LUMS}\\[2pt]
{\small\texttt{\{yixi.chen, muhammad.bilal\}@kaust.edu.sa}}}
\date{}

\begin{document}
\pagestyle{plain}
\maketitle
\begin{strip}
\centering
\begin{minipage}{0.8\textwidth}

\itshape``The Semantic Web will bring structure to the meaningful
content of Web pages, creating an environment where software agents
roaming from page to page can readily carry out sophisticated tasks
for users.''\\
\hspace*{\fill}--- Berners-Lee, Hendler, and Lassila,
\emph{The Semantic Web}, 2001~\cite{lassila2001semantic}\par
\vspace{4pt}\normalfont
Twenty-five years later, the AI agents have arrived. The structure has not.

\end{minipage}
\end{strip}
\begin{abstract}
AI agents, not humans, are becoming the web's primary consumers, yet
the web still ships content structured for human browsing, with no
native way for publishers to expose the semantic units within it.
Access is therefore mediated by a handful of AI providers that crawl,
chunk, and index content unilaterally, leaving publishers little
control over how their content is represented, retrieved, or
attributed. We argue this is a structural problem, not one better
chunking or retrieval algorithms can solve. The web lacks a native
layer for publisher-controlled semantic discovery and retrieval.
We propose \proposed, an \ainet architecture that supplies this
layer. Publishers expose semantically meaningful units (e.g., passages,
claims, tables) through explicit, policy-governed interfaces they
control, and a web-native semantic resolver, analogous to DNS but
operating over meaning rather than location, lets agents discover
relevant sources before retrieving fine-grained content directly
from publishers. Our end-to-end prototype matches or exceeds representative centralized
retrieval pipelines in answer quality, with up to $4.2\times$ higher
accuracy on large, specialized document collections. Against a public
web-search agent, it answers up to $4.2\times$ more questions
correctly, while moving $2.9$--$41\times$ less data per query. 
Because retrieval terminates at
publisher-operated servers, attribution, provenance, and usage
policies travel with the content, realigning incentives across the
web ecosystem.\footnote{Data: \url{https://github.com/sands-lab/aistral-dataset}}
\end{abstract}

\section{Introduction}
\label{sec:intro}

In 2001, Tim Berners-Lee and colleagues sketched the web's next act:
a Semantic Web whose content would carry machine-readable meaning, letting software agents traverse pages and complete tasks on their users' behalf~\cite{lassila2001semantic}. The vision was right about the destination and early about the vehicle. Its intended consumers were rule engines that could not interpret unstructured content, so meaning had to be formalized in advance,
in standards such as RDF and OWL, and interoperation required
agreeing on shared vocabularies up front. Publishers never took on
that burden at scale, and the effort stalled, as its own authors
later chronicled~\cite{shadbolt2006semantic}.

A quarter-century later, the machine consumers have arrived, just not the ones the Semantic Web was designed for. Rather than requiring publishers to encode meaning in a predefined ontology, LLM-based agents infer it from ordinary, human-oriented content as they answer a query, and they now dominate consumption of the web: as of August 2026, bots account for 60.8\% of Internet traffic~\cite{cloudflareradar2026}, with AI scraper traffic alone growing 597\% over the past year~\cite{humansecurity2026}.
But because the web still exposes primarily its human-facing representation, a version of the machine-readable web is being built anyway by AI providers.
AI search systems construct machine-usable representations through a mix of provider-controlled crawling and query-time access~\cite{webconf26}.
Proprietary retrieval pipelines fetch pages, chunk them (split them into retrieval-sized passages), embed them (map them to vectors for similarity search), and store them in searchable indexes~\cite{anthropic2024contextual, tavily_aws2026, linkup_search2026}.

The economics are starkly asymmetric. Traditional search kept crawling and referral loosely coupled (Google's crawl-to-refer ratio is roughly 5.2:1), whereas frontier AI companies crawl over a thousand pages for every visitor they refer back~\cite{cloudflareradar_crawltorefer2026}, and cite only a small fraction of the pages they access when generating an answer~\cite{webconf26}.\footnote{For example, the study~\cite{webconf26} finds that the GPT-5 Thinking model cites only one URL for approximately every five webpages it accesses.}
Publishers watch zero-click interactions erode the value their content creates~\cite{bain2024zeroClick}, armed only with coarse, crawler-identity-level controls such as \texttt{robots.txt} and \texttt{llms.txt}~\cite{cloudflare_contentindependence2025}. Meanwhile, which sources are consulted, cited, or ignored is increasingly decided by a few providers~\cite{guardian_ai_journalism_coalition_2026,nyt_perplexity_lawsuit_2025,ft_ai_scraping_publishers_2026,cloudflare_ai_crawler_permission_2025}, introducing a new layer of centralized gatekeeping atop an originally decentralized web.

We argue these are two failures of the same missing layer. The Semantic Web asked publishers to do too much: formalize meaning up front, coupling how content is \emph{represented} to how every consumer would \emph{interpret} it, and it collapsed under that coordination burden. Today's AI-mediated web asks publishers for nothing and takes everything: providers reverse-engineer semantic units from human-oriented pages lossily and unilaterally. At the same time publishers have no interface through which to express control at all. What both lack is a native web layer for publisher-controlled semantic discovery and retrieval; recent work frames AI agents as an infrastructure problem~\cite{chan2025infrastructure}, and this, we argue, is the missing infrastructure. 

Crucially, the consumer shift is what finally makes such a layer feasible: because LLM-based agents can interpret ordinary content without a predefined schema, publishers need only \emph{expose} their content's semantic units, not formalize their meaning for the entire world.
Emerging agent protocols do not provide this layer. MCP~\cite{anthropic2024mcp} (\autoref{sec:comm-proto}) standardizes how an agent talks to a
server it already knows, much as TCP standardizes delivery between
two known endpoints. It leaves open what publishers expose, and how
uncoordinated publishers become uniformly discoverable and
retrievable by \emph{any} agent, the same naming-and-semantics gap
HTTP filled over TCP. 

\begin{figure}[t]
    \centering
    \includegraphics[width=\columnwidth]{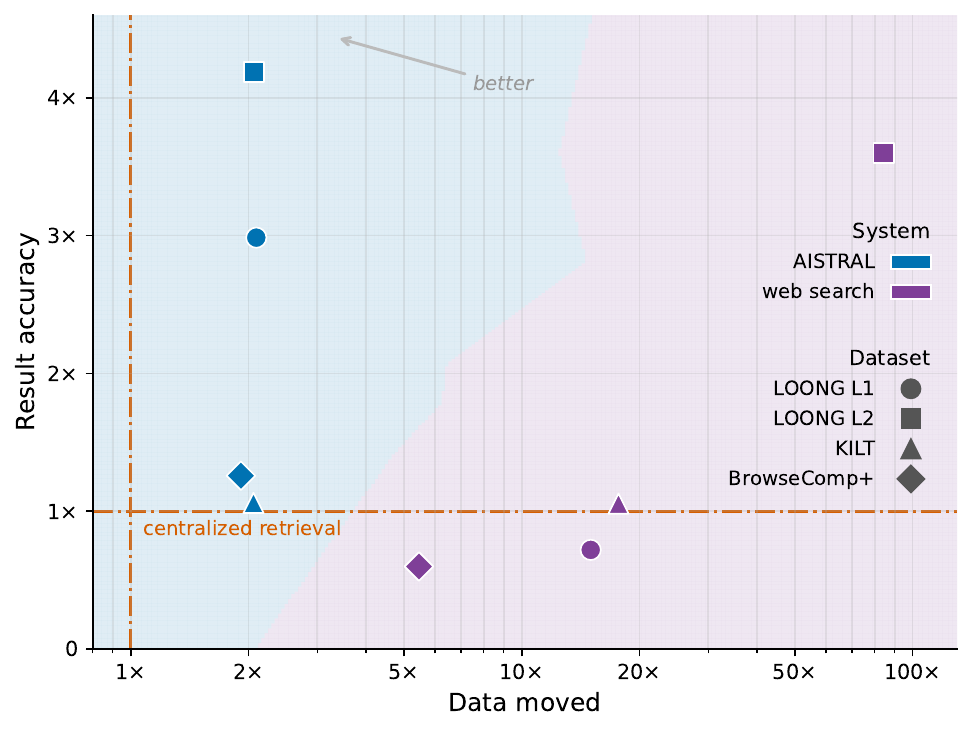}
  \caption{Accuracy versus data cost, each as a multiple of the baseline (centralized retrieval, dashed lines at $1\times$).
    Data moved per query on the horizontal axis and answer accuracy on the vertical. 
    Each system has one point per benchmark (marker shape),
    spanning multiple domains and question types, from encyclopedic multi-hop questions (KILT-HotpotQA) and obfuscated public-web puzzles (BrowseComp+) to single-fact lookups and multi-document
    aggregations over financial filings (LOONG~L1/L2). 
    \proposed (blue) stays near the baseline's data cost ($\mathord{\sim}2\times$) while
    $1.1$--$4.2\times$ more accurate, whereas the web-search agent (purple) moves $5$--$80\times$ more data for mixed accuracy.
    Up and to the left is better.\protect\footnotemark{}
   }
    \label{fig:tradeoff-scatter}
\end{figure}
\footnotetext{The \emph{web-search agent} is an off-the-shelf LLM (\texttt{gpt-5-mini}) that answers each query with live web search instead of our indexed corpus (\autoref{sec:eval}).}

This paper fills this gap with \proposed, a hierarchical, publisher-first architecture that standardizes roles and interfaces
for a semantic web at Internet scale. It completes, we argue, the Semantic Web's journey with its coupling inverted: meaning stays with publishers, and interpretation moves to agents.
Concretely, the \proposed \ainet is characterized by four core design principles:

\begin{smartlist}
\item \textbf{Publisher-Controlled Semantic Interfaces:} Publishers expose queryable endpoints for the semantic units, such as passages, claims, tables, and multimodal segments, together with provenance, attribution, licensing, and access-policy metadata. The interfaces may be implemented over MCP, but \proposed adds what MCP leaves undefined: what publishers expose, and how uncoordinated publishers become discoverable. %

\item \textbf{Semantic Discovery:} A web-native semantic resolver maps an agent's query and constraints to relevant publisher-operated endpoints using lightweight source-level records. It returns source references, not content or answers, separating source selection from content access so that no single provider controls both stages end to end.

\item \textbf{First-Class Semantic Retrieval:} Retrieval of semantic units becomes a native web abstraction. Given the endpoints discovery returns, an agent queries each publisher directly and receives the units relevant to the query, without downloading or parsing full documents. 
\item \textbf{Internet-Scale Efficiency:} The architecture must be built to scale to the size of the web itself, not adapted to it afterwards.
\end{smartlist}

These principles are realized by three independently owned roles connected by three narrow interfaces.
\begin{smartlist}
\item \emph{Publishers} run \textbf{Agentic Twins}: servers under publisher control that expose a source's content as semantic units rather than as pages to be scraped, populated through a \texttt{publish} interface.
\item \emph{Infrastructure providers} run \textbf{semantic resolvers}, which operate analogously to DNS but over meaning rather than location: Agentic Twins register lightweight discovery records (per-source summaries and metadata, not content) through \texttt{insert}, and an AI agent's \texttt{resolve} query returns the addresses of relevant Agentic Twins, not answers.
\item The \emph{AI agent} (the client) then invokes \texttt{search} against the selected Twins to retrieve fine-grained semantic units directly from publishers, and synthesizes the answer locally.
\end{smartlist}

Discovery is thereby separated from retrieval: the resolver never holds content, and retrieval always terminates at a publisher-operated server. This hierarchy contrasts sharply with today's centralized retrieval pipelines: instead of a single entity deciding how content is chunked and retrieved, responsibility is distributed across publishers, semantic resolvers, and AI agents, with clear interfaces and control boundaries (\autoref{fig:aistral-architecture}).

Beyond efficiency, the architecture realigns incentives across the web
ecosystem. Publishers keep control over how their content is accessed,
exposing semantic units through Agentic Twins they operate, while the
resolver indexes only per-source discovery metadata
(\autoref{fig:aistral-architecture}). Attribution, provenance, and
usage-policy fields travel with each unit from publication to final
retrieval, rather than being stripped out during centralized crawling
and re-indexing (\autoref{sec:discussion}). AI developers gain cleaner,
lower-latency access to structured knowledge without brittle scraping
pipelines, infrastructure providers shed redundant crawler-induced
load, and users receive more transparent, attributable responses.
Together, these incentives suggest an adoption path built on the web's
decentralized governance rather than on unilateral cooperation from
dominant AI platforms. Early industry motion supports this path:
Cloudflare now builds per-publisher AI indexes with an aggregated
cross-site index~\cite{cloudflare_aiindex2025} and meters publisher
compensation per crawl~\cite{cloudflare_paypercrawl2025}, though as
single-vendor products rather than the open, interoperable layer we
argue for (\autoref{sec:discussion}).

We implement and evaluate \proposed end-to-end against realistic
baselines. \autoref{fig:tradeoff-scatter} summarizes the result:
\proposed stays near centralized retrieval's own data cost
($\sim2\times$) while matching or exceeding its accuracy, and clearly
outperforms a public web-search agent, which needs $5$--$80\times$
more data for mixed accuracy (\autoref{sec:eval}).

The Semantic Web's goal, a web AI agents can consume with publisher
intent intact, no longer requires global ontological consensus. It
requires \textit{rethinking} who does the work of preparing content
for agents, and who controls how that content is accessed.
Publishers expose, resolvers discover, agents interpret, and
interfaces make that division binding. Agents already approximate the
applications the Semantic Web promised, composing answers from many
independent sources, but they do so by scraping, without provenance,
permission, or payment, and least reliably on the specialized sources
where open-web search falls short (\autoref{fig:tradeoff-scatter}).
With such a layer, the same applications could retrieve each unit
directly from its publisher, with provenance and usage terms attached
and each source cited and compensated, sustaining the publishing they
depend on rather than eroding it. As a vision paper grounded in a
concrete system design, this work takes a first step. We introduce a
publisher-first architecture, implement multiple semantic resolver
designs, and evaluate them end-to-end. Challenges remain, including
resolver governance, incentive-compatible deployment, trust and
ranking, and retrieval and caching optimization
(\autoref{sec:discussion}). We hope this work catalyzes a broader
community effort toward a web that AI agents read and its publishers,
at last, get to write.

\section{Current AI-based Web search}
\label{sec:current-research}

In its simplest form, current AI-based web search works as follows. 
When a user submits a query, the AI system first performs a web search using either external search engines (e.g., Google, Bing) or a proprietary index maintained by the AI provider to identify the most relevant, high-quality sources. 
The system then accesses the web pages, parses each page into an ingestible format, chunks the content, and stores it in a vector database.
Deployed systems already work this way: Tavily, for example, crawls web content once and serves it from its own cache for every later query~\cite{tavily_aws2026}.

The natural language query is translated into a vector embedding using an embedding model. 
This is used to retrieve relevant chunks from the vector database, and an LLM generates its answer based on this in-context information. 
This process is increasingly ``agentic,'' meaning the system may decompose complex questions into sub-tasks, search for each separately, and aggregate the results. 
If information is insufficient or ambiguity remains, the system iterates over previous steps. 
Additionally, knowledge graphs and keyword-based search might be incorporated for further improvement. 
To ensure accuracy, these systems ground responses with inline citations, enabling users to verify facts against original sources.

While this centralized architecture offers a global view of information and corresponding benefits, it has significant structural drawbacks, as depicted in ~\autoref{fig:tradeoff-scatter}. 
Because a single provider owns crawling, chunking, embedding, and indexing end-to-end, publishers have no visibility into how their content is represented internally and no means to influence it.
That is, the provider alone decides every re-indexing choice, from adopting a new embedding model to changing the chunking policy for the whole corpus, no matter the compute and storage cost. 
This also means chunking and embedding choices stay fixed centrally, the same choices applied to every publisher's content.
Yet prior work shows the opposite is true: the best chunk size and embedding model depend on the data and domain~\cite{yepes2024financialreportchunking, bhat2025rethinkingchunksizelongdocument, tang2025needdomainspecificembeddingmodels}.
This suggests that content-aware choices, made per publisher instead of once for the whole web, would perform better.
The result is an architecture that blocks publisher control by design, not just by making it inconvenient: publishers have no interface to express a preference, even if the provider wanted to honor one.

There is a deeper issue underneath this: the content being chunked and embedded was never written for that purpose.
Webpages are built for people to read, so they rely on narrative pacing, visual emphasis, and figures that only make sense in context.
So crawling, chunking, and embedding systems have to work backward from this human-focused format, trying to recover the meaningful units and facts it contains.
This process is inherently lossy and fragile, no matter how much compute or engineering effort goes into it.

We use this centralized AI-based status quo web search design as our baseline for comparison against the \proposed \ainet architecture.

\subsection{Related Work}
\label{sec:related-work}

\subsubsection{Communication Protocols.}
\label{sec:comm-proto}
Model Context Protocol (MCP)~\cite{anthropic2024mcp} and
Agent-to-Agent (A2A)~\cite{google2025a2a} standardize how two parties that already know each other exchange messages, much as TCP standardizes delivery between two endpoints that already know how to
interpret each other's bytes. Neither addresses how independent, uncoordinated publishers expose content so that \emph{any} agent can
uniformly discover and retrieve it. This is the same
naming-and-semantics gap HTTP filled over TCP.

This gap spans both halves of \proposed.
An Agentic Web Server is not merely an MCP server, but a role that standardizes how a publisher prepares and exposes semantic units in the first place, a question MCP leaves entirely to each server's implementation.
Also, a semantic resolver is not merely a sharper MCP registry, since the registry catalogs static, developer-published servers.
It has no convention for a publisher's dynamic corpus changes or a query narrowing that catalog by semantic relevance. These are the two
halves of the contract \proposed standardizes (\autoref{sec:ainative}).

\subsubsection{Semantic Web.} 
The Semantic Web~\cite{lassila2001semantic} envisioned a web of machine-readable data through ontologies and formal metadata (RDF, OWL), later revisited by its own authors as progress stalled at web scale~\cite{shadbolt2006semantic}.
At its core, it assumes meanings among all documents must be formalized and agreed upon in advance, coupling how content is \emph{represented} to how it is \emph{interpreted}.
This required global ontological consensus that proved difficult to reach, and is largely orthogonal to LLM-based agents, which interpret content flexibly at retrieval time without a predefined schema.
SOLID~\cite{solid} pursues a related decentralization goal but from a different angle, letting individuals own and control their personal data in self-hosted pods rather than platform silos.
However, it only addresses data ownership, not publisher-side content discovery or retrieval for AI agents.

Closer to our setting, NLWeb~\cite{nlweb} lets any website expose an MCP-backed natural-language query endpoint over its own content, reusing existing Schema.org markup as a lightweight semantic substrate\footnote{This is structurally similar to an Agentic Twin's \texttt{search()} in \proposed.}.
However, NLWeb provides no equivalent discovery layer.
It means an agent \emph{must} already know which site to query.
Without such a discovery layer, it has to query every candidate site one by one, introducing exactly the uncontrolled, site-by-site search that we try to avoid.
\proposed's semantic resolver is precisely this missing piece: decoupling representation from interpretation as LLMs enable, while still bounding discovery at web scale.

\subsubsection{Storage and Indexing.} Separately, work on storage systems targets the scale and latency demands of AI-driven retrieval.
Modern vector databases demonstrate how chunk-level semantic indexing can scale to billions of items~\cite{vectorbenchmark, pan2024survey, jayaram2019diskann}, providing the infrastructure that retrieval pipelines build upon.
These works optimize \emph{how} an index is stored and queried, independent of \emph{what} should be indexed or \emph{who} controls that decision.

\subsubsection{Retrieval-Technique Optimization.} A large and active literature improves retrieval quality within a fixed pipeline.
Chunking strategy is known to significantly affect retrieval quality, with optimal chunk size and boundary split varying by document type and domain~\cite{bhat2025rethinkingchunksizelongdocument, yepes2024financialreportchunking}.
Embedding choice is similarly domain-sensitive, motivating works on domain-specific and task-specific embedding models~\cite{tang2025needdomainspecificembeddingmodels}.
Beyond flat vector similarity, knowledge-graph-augmented retrieval (GraphRAG~\cite{han2024graphrag}) captures entity and relational structure that pure embedding similarity can miss~\cite{edge2024graphrag, han2025ragvsgrag}.
Each of these techniques improves retrieval quality within a single, centrally-controlled pipeline. 
None of them ask \emph{who} controls it, \emph{how} content enters it, or \emph{how} it stays current as sources change.
\proposed instead exposes explicit interfaces (\autoref{appen:endpoint-datamodels}), giving publishers and resolvers a defined place to handle these questions.

\subsubsection{Semantic Caching.} A related line of work targets serving efficiency rather than retrieval quality directly, caching semantically similar queries or sub-results to reduce redundant computation and latency~\cite{gill2025advancing, ruan2025cortex, wang2025categoryaware}.
This is largely orthogonal to our contribution: caching can be layered atop any component, as we note in \autoref{sec:discussion}.

However, each of these efforts is ad-hoc, optimizing one piece of the pipeline in isolation: agent-tool communication, index storage, retrieval quality, or serving latency. 
They treat the pipeline's overall structure and ownership as fixed. 
Yet collaborating across components can unlock gains neither could reach alone, as shown by algorithm-system co-design work combining retrieval and generation~\cite{jiang2024piperag}.
\proposed instead targets the boundaries \emph{between} these components: \emph{who} controls content representation, \emph{how} sources are discovered, and \emph{how} retrieval responsibility is distributed across publishers, resolvers, and agents.
We view this as complementary rather than competing: \proposed builds on and can incorporate all of the above: an agentic web server can use MCP to expose tools, a vector database to index its content, GraphRAG to structure it, and semantic caching to serve efficiently.
In addition, \proposed provides the missing structural layer that determines who makes these choices, and how the results become discoverable across the open web.

\section{System Architecture}
\label{sec:ainative}

\subsection{Design Objectives}

We introduce the concept of an \ainet, an architectural evolution of the web designed to natively support semantic retrieval, agentic workflows, and fine-grained information access. Rather than treating the web as a collection of HTML documents to crawl and parse, the AI-Native Internet exposes structured, semantically indexed interfaces optimized for LLM-based AI systems.
The key design objectives for \ainet are as follows:

\begin{smartlist}
    \item \textbf{Publisher-First Control and Incentive Alignment (O1):}  
    Website operators should retain explicit control over how their content is accessed by AI systems. 
    This includes deciding what information is exposed, how content is structured for AI consumption, how data is chunked and where provenance, attribution, copyright signals, and usage policies attach as metadata carried through the interfaces.
    By shifting these decisions from centralized AI providers to publishers, the architecture enables clearer incentive alignment across publishers, AI developers, infrastructure providers, and users.

    \item \textbf{Semantic Discovery (O2):}  
    Semantic discovery should be hierarchical rather than centralized. 
    AI agents should first discover relevant information sources through a web-native semantic resolver before retrieving fine-grained content directly from publishers. 
    This separation of source discovery from content access avoids opaque, end-to-end ranking pipelines and distributes responsibility across semantic resolvers, publishers, and agents, mirroring the Internet’s existing hierarchical design principles (e.g., DNS).

    \item \textbf{AI-Native, Semantic-First Access (O3):}  
    As discussed in~\autoref{sec:current-research}, today's web content is authored for human cognition, forcing AI systems to reverse-engineer semantically meaningful units (e.g., passages, claims, tables, multimodal segments) from content built for human readers. 
    The architecture should instead treat these semantic units as a first-class, publisher-exposed abstraction with a native place to live, rather than something every consumer must independently re-derive.

    \item \textbf{Efficiency and Scalability (O4):}  
    As LLM-based AI agents issue increasingly complex, multi-step queries, the architecture must scale efficiently. 
    By enabling scoped retrieval of relevant sources and fine-grained content, the system should minimize redundant crawling, excessive bandwidth usage, and repeated re-indexing, while remaining practical to deploy and operate at Internet scale.
\end{smartlist}

Together, these design objectives inform \proposed, a publisher-first architecture in which semantic structure and access control are embedded into the web’s content-publishing and retrieval substrate, rather than concentrated within a small number of centralized AI platforms.

\subsection{System Overview}
\label{sec:system-overview}

\begin{figure*}[t]
\centering

\begin{tikzpicture}[x=1cm,y=1cm,>=Stealth,font=\small]

\tikzset{
  mainbox/.style={draw=aisGray,rounded corners=3pt,line width=0.8pt,fill=white,align=center,minimum height=1.0cm},
  sourcebox/.style={mainbox,minimum width=2.55cm},
  serverbox/.style={draw=aisOrange,rounded corners=3pt,line width=0.9pt,fill=aisOrange!9,align=center,minimum width=3.25cm,minimum height=1.35cm},
  resolverbox/.style={draw=aisGreen,rounded corners=5pt,line width=0.9pt,fill=aisGreen!9,align=center,minimum width=3.25cm,minimum height=1.05cm},
  retrieverbox/.style={draw=aisBlue,rounded corners=3pt,line width=0.9pt,fill=aisBlue!8,align=center},
  readerbox/.style={draw=aisGray,rounded corners=3pt,line width=0.8pt,fill=aisGray!6,align=center},
  outerbox/.style={draw=aisGray!55,rounded corners=4pt,line width=0.6pt,dash pattern=on 3pt off 2pt},
  userbox/.style={mainbox,minimum width=1.9cm},
  offarrow/.style={->,draw=aisOrange,line width=1.25pt,dashed},
  onarrow/.style={->,draw=aisBlue,line width=1.35pt},
  responsearrow/.style={->,draw=aisBlue!70,line width=1.0pt},
  offstep/.style={circle,draw=aisOrange,fill=aisOrange!15,line width=0.8pt,inner sep=0.9pt,font=\tiny\bfseries},
  onstep/.style={circle,draw=aisBlue,fill=aisBlue!12,line width=0.8pt,inner sep=0.9pt,font=\tiny\bfseries},
  offlabel/.style={fill=white,text=aisOrange,font=\scriptsize\bfseries,inner sep=1.2pt},
  onlabel/.style={fill=white,text=aisBlue,font=\scriptsize\bfseries,inner sep=1.2pt}
}

\fill[aisBlue!5,rounded corners=7pt]   (-1.0,-1.15) rectangle (6.65,1.75);   %
\fill[aisGreen!5,rounded corners=7pt]  (7.05,1.25)  rectangle (11.90,3.15);  %
\fill[aisOrange!5,rounded corners=7pt] (7.05,-1.15) rectangle (15.20,1.15);  %

\node[font=\bfseries,text=aisBlue]   at (2.82,1.42) {Client};
\node[font=\bfseries,text=aisGreen]  at (9.35,2.88) {Infrastructure Provider};
\node[font=\bfseries,text=aisOrange] at (13.65,0.83) {Publisher};

\node[sourcebox] (source) at (13.65,-0.08) {Content\\Source};

\node[serverbox] at (9.35, 0.12) {};
\node[serverbox] at (9.45, 0.02) {};
\node[serverbox] (server) at (9.55,-0.08)
  {Agentic Twin(s)\\[2pt]{\scriptsize Semantic Units}};

\node[resolverbox] (resolver) at (9.55,2.01)
  {Semantic Resolver\\[2pt]{\scriptsize Source Index}};

\node[userbox] (user) at (0.35,-0.12) {User};

\draw[outerbox] (2.45,-0.90) rectangle (6.25,1.07);
\node[font=\bfseries,text=aisGray] at (4.35,0.78) {AI Agent};

\draw[retrieverbox] (2.55,-0.02) rectangle (6.15,0.62);
\node[anchor=west,font=\bfseries\scriptsize,text=aisBlue] at (3.08,0.45) {Retriever};
\node[onstep] at (2.85,0.28) {Q1};
\node[anchor=west,font=\scriptsize\ttfamily] at (3.12,0.18) {process\_query()};

\draw[readerbox] (2.55,-0.68) rectangle (6.15,-0.08);
\node[anchor=west,font=\bfseries\scriptsize,text=aisGray] at (3.08,-0.26) {Synthesizer};
\node[onstep] at (2.85,-0.40) {Q4};
\node[anchor=west,font=\scriptsize\ttfamily] at (3.12,-0.50) {synthesize()};

\draw[offarrow] (source.west) -- (server.east);
\node[offlabel] at (11.82,0.13) {publish()};
\node[offstep]  at (11.82,-0.22) {P1};

\draw[offarrow] (server.north) -- (resolver.south);
\node[offlabel] at (10.10,1.04) {insert()};
\node[offstep]  at (9.13,1.04) {P2};

\draw[onarrow]       (4.85,0.62) -- (7.92,2.31);
\draw[responsearrow] (7.92,2.01) -- (5.46,0.62);
\node[onlabel] at (6.23,1.8) {resolve()};
\node[onstep]  at (6.28,1.48) {Q2};
\node[font=\scriptsize,text=aisBlue,align=center] at (7.00,1.05) {retr.\\idx};

\draw[onarrow]       (6.13,0.22)  -- (7.92,0.22);
\node[onlabel] at (6.98,0.44) {search()};
\node[onstep]  at (6.98,0.1) {Q3};
\draw[responsearrow] (7.92,-0.35) -- (6.13,-0.35);
\node[font=\scriptsize,text=aisBlue,align=center] at (7.00,-0.65) {retr. \\[-2pt]chunks};

\draw[responsearrow] (1.31,0.20) -- (2.57,0.20);
\node[font=\scriptsize,text=aisBlue] at (1.84,0.42) {submit};

\draw[responsearrow] (2.57,-0.35) -- (1.31,-0.35);
\node[font=\scriptsize,text=aisBlue] at (2.04,-0.55) {return};

\end{tikzpicture}
\caption{\proposed architecture and interfaces across publisher, infrastructure-provider, and client roles.
Orange dashed arrows denote publication and ingestion, while blue solid arrows denote query-time interactions.
The Client's AI agent is decomposed into a \textbf{Retriever} (query understanding and source discovery/retrieval) and a \textbf{Synthesizer} (context synthesis).}
\label{fig:aistral-architecture}
\end{figure*}

\autoref{fig:aistral-architecture} presents the high-level architecture of \proposed achieving \ainet. 
Rather than relying on a centralized retrieval service, \proposed separates content publication from query resolution through two complementary workflows.

The publication workflow allows publishers to expose semantic content while retaining full control over content preparation and hosting (\textbf{O1}). 
Publishers publish their content through agentic web servers, which perform semantic preprocessing (e.g., chunking and embedding) and register source-level metadata with the semantic resolver\footnote{Throughout this paper, \proposed is illustrated using a deployment model where infrastructure providers host the agentic web servers.
The architecture itself does not depend on this particular deployment, and alternative ownership models are discussed in \autoref{sec:discussion}.}. 
We call each such server an \emph{Agentic Twin}: a publisher-controlled, queryable counterpart to a website, exposing the site's semantic units through the \texttt{search} interface (\autoref{sec:query-resolution}) rather than serving pages for human browsing.
This establishes a lightweight discovery layer without centralizing the underlying content.

The query-resolution workflow begins when an AI agent receives a user query. 
Internally, the agent is organized into two components: a \textbf{retriever}, which performs source discovery through the semantic resolver to identify relevant agentic web servers and then retrieves semantically relevant content directly from those sources, and a \textbf{synthesizer}, which synthesizes the retrieved content into a final response. 
This work's contributions center on the retriever; the synthesizer follows a standard retrieval-augmented generation design.
By separating source discovery from semantic retrieval, \proposed preserves publisher autonomy while enabling scalable and decentralized knowledge access.

\subsection{Semantic Publication}
\label{sec:semantic-pub}

The semantic publication workflow enables publishers to expose AI-consumable content while retaining ownership over content preparation, hosting, and access policies. 
Rather than uploading documents to a centralized index or exposing them to crawlers, publishers make semantic endpoints discoverable through lightweight source registration.

\paragraph{Semantic Preparation.}
The publication workflow begins at \Circled{P1}, where publishers expose AI-consumable content through the \texttt{publish} interface on an agentic web server (\textbf{O3}). 
Rather than prescribing a fixed semantic preprocessing pipeline, \proposed deliberately delegates semantic preparation to content publishers, according to application requirements. 
This preparation may include document chunking, embedding generation, metadata extraction, or alternative semantic representations suitable for downstream retrieval. 
By decoupling semantic preparation from the architecture itself, \proposed remains agnostic to application domains and future retrieval techniques while allowing publishers to retain full control over how their content is represented and maintained.

\paragraph{Source Registration.}
After semantic preparation, \Circled{P2} registers the published source with the semantic resolver through the \texttt{insert} interface. 
The resolver does not replicate publishers' full prepared content or fine-grained semantic units; instead, it stores a lightweight snippet sufficient for discovery (e.g., a short content summary per source) alongside source identifiers, endpoint locations, and optional metadata.
\texttt{Resolve} matches queries against this snippet to determine relevance, while metadata serves a different role, narrowing candidates by constraints such as freshness, domain, or policy (\autoref{sec:discussion}).
Consequently, the resolver functions as a lightweight discovery layer, while the underlying knowledge remains hosted and managed by the corresponding publishers.

The registered sources subsequently become available for discovery during the query-resolution workflow, described in the following subsection.

\subsection{Query Resolution}
\label{sec:query-resolution}
The query-resolution workflow transforms user queries into grounded responses by progressively acquiring knowledge from decentralized semantic sources. 
Unlike conventional retrieval architectures that couple source discovery and content retrieval within a centralized service, \proposed explicitly separates these two operations (\textbf{O2}). 
This separation enables AI agents to dynamically discover relevant semantic sources before retrieving knowledge directly from the corresponding agentic web servers.

\paragraph{Query Understanding.}
The workflow begins where the AI agent submits a user query. 
During \Circled{Q1}, the \textbf{retriever} interprets the query according to the application's objectives and execution strategy. 
This stage may involve query rewriting, decomposition, planning, or other reasoning mechanisms prior to external knowledge acquisition.

Query decomposition itself is not a new concept.
Major AI providers already do this.
What \proposed proposes is \emph{where} it happens.
Today, decomposition happens inside the centralized AI provider, together with crawling and indexing.
In \proposed, it runs on the client, as part of the retriever.
No single party needs to own both decomposition and content access anymore.

\paragraph{Source Discovery and Semantic Retrieval.}
Following query understanding, the \textbf{retriever} performs two sequential operations. 
At \Circled{Q2}, it queries the semantic resolver to identify relevant semantic sources and obtain the corresponding retrieval endpoints. 
The retriever then invokes \Circled{Q3} to retrieve semantically relevant information directly from the discovered agentic web servers. 
This separation enables the resolver to operate solely as a discovery service, while semantic retrieval remains fully decentralized.

\paragraph{Response Generation.}
Once the required information has been collected, the \textbf{synthesizer} aggregates the retrieved data at \Circled{Q4}, synthesizing the final response. 
Finally, the agent returns the generated response to the user. 
Because retrieval is performed directly against the discovered semantic sources, the synthesized response reflects the most up-to-date publisher-provided knowledge available during execution.

The full specifications of these interfaces, together with the implementations of the semantic resolver and their ingestion and query-resolution behavior, are provided in~\autoref{appen:endpoint-datamodels}. 

\section{Evaluation}
\label{sec:eval}

Our evaluation answers four questions:
\begin{smartlist}
    \item \textbf{Does the architecture hold up?}
    Does separating source discovery from content retrieval reach the same answer quality as a centralized status quo pipeline that owns both?
    \item \textbf{What does decentralization cost or save?}
    How much data moves per query under \proposed, compared to a centralized retriever and a web-search agent (\textbf{O4})?
    \item \textbf{Where do errors come from?}
    When the pipeline gets an answer wrong, is the cause routing, retrieval, or synthesis?
    \item \textbf{Does this generalize?}
    Do the answers to (1)-(3) still hold across different kinds of corpora, either small and diverse, large and repetitive, or adversarial public-web content?
\end{smartlist}

\begin{figure*}[t!]
    \centering
    \includegraphics[width=0.95\linewidth]{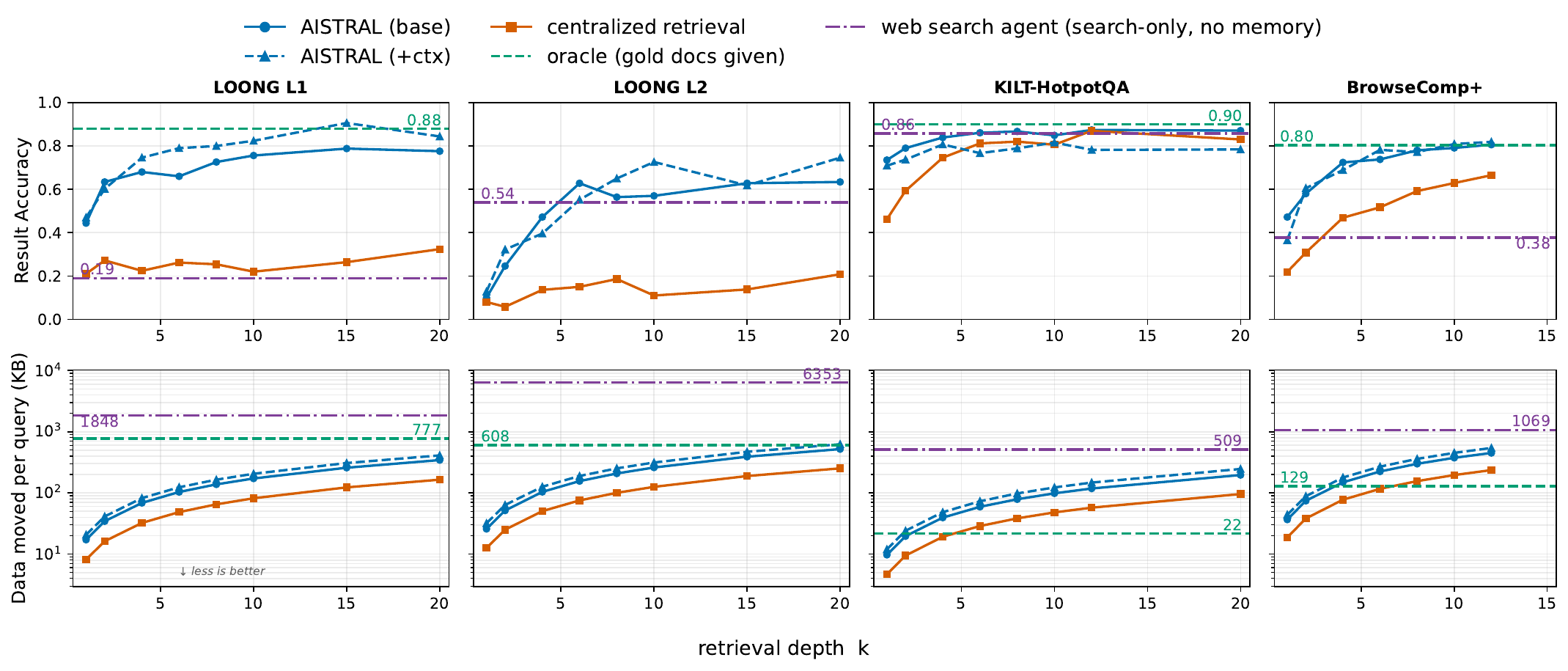}
    \caption{\textbf{Answer correctness vs.\ data moved, across four benchmarks.}
    Columns are datasets. 
    Top row: answer correctness. 
    Bottom row: data transferred per query to produce that answer (KB, log scale, shared across panels; lower is better). 
    Both are plotted against retrieval depth $k$ (chunks per agentic twin).
    Two $k$-independent baselines appear as horizontal lines in \textbf{both} rows: \textbf{perfect document retrieval} (green --- the reader given the full gold documents) and a \textbf{web search agent (search-only, no memory)} (purple --- the same reader answering from OpenAI's live public-web search). 
    \proposed approaches the perfect-retrieval ceiling and stays well above the web agent on specialized corpora (LOONG, BrowseComp+)\protect\footnotemark{}, while moving far less data than the web agent.}
    \label{fig:tradeoff-k-sweep}
\end{figure*}
\footnotetext{BrowseComp+ is shown only to $k{=}12$: at larger $k$ the retrieved context reaches 0.8--2\,MB per query, exceeding the reader model's input limit, and ${\sim}11\%$ of queries return an empty answer.}

\para{Experimental Setup:}
We run experiments on an Ubuntu 22.04.5 LTS server running in a QEMU virtualized environment, with 32 vCPUs (AMD EPYC 9754) and 64 GB RAM.
We ingest each \emph{unique} document exactly once and split it into passages using LangChain's \textsf{RecursiveCharacterTextSplitter}~\cite{langchain-splitter} with a chunk size of 1{,}500 characters and an overlap of 150 characters; we selected this chunk size via a sweep, as smaller chunks degrade retrieval by fragmenting the statement- and table-level context that queries match against, consistent with prior findings that optimal chunk size depends on document type and content structure~\cite{bhat2025rethinkingchunksizelongdocument, yepes2024financialreportchunking}.
Passages are embedded with OpenAI's \textsf{text-embedding-3-small} model, whose unit-normalized vectors make Euclidean ($L_2$) and cosine ranking equivalent, and are indexed in a per-server \textsc{ChromaDB} collection.
Documents are placed by \emph{source identity}: the filing entity (company) for LOONG, or the source web domain for the URL-grounded datasets.
Each identity is hashed to a single server, so all documents sharing an identity live on the same server\footnote{We use SHA-256 over the identity key. We assume one content provider hosts all documents under a given identity, and that identities hash roughly uniformly across the $N$ servers.}.
A multi-hop query draws on several distinct identities (different companies or different web domains).
Its evidence therefore lands on different servers, and answering it requires routing across them.

Complicated questions requiring multi-hop reasoning are decomposed into a fixed set of sub-queries that are routed independently, executed in parallel, and finally have results merged.
The decomposition is single shot, so a sub-query cannot be conditioned on entities resolved by another\footnote{We do not handle ``bridge'' questions, where one hop finds an entity the next hop looks up. 
This is out of scope: our contribution is the structure, not retrieval optimization. We leave it to future work.}.
We deploy 10 agentic twin servers.
For each sub-query, the resolver returns the semantically relevant twins (up to half of all twins) and we retrieve the top-$k$ chunks from each.
We sweep $k$ (the per-server retrieval depth) in our experiments. The passages retrieved across servers are concatenated into a single context, from which \textsf{gpt-5-mini} synthesizes the final answer, grounded strictly on the retrieved context and not from its own internal knowledge (\autoref{app:prompts}).
A query is scored $0$ and counted as a failure when no server returns any passage (empty context), or when the synthesizer reports the retrieved passages lack the answer (irrelevant context).

\para{Dataset:} We want benchmarks that stress a decentralized resolver in different ways: \emph{how many documents a single query needs}, \emph{how topically diverse or near-duplicate the corpus is}, and \emph{how queries are phrased}.
We also want to attribute each failure to a specific stage (i.e., routing, chunk retrieval, or synthesis) rather than only scoring the final answer.
The second goal requires close-ended answers, document-level gold labels, and fine-grained evidence that is localized to a few spans\footnote{We also evaluate with an open-ended dataset in \autoref{app:open-end-eval}.}.
Three benchmarks meet the requirements: KILT-HotpotQA~\cite{petroni-etal-2021-kilt}, LOONG~\cite{loong2025}, and BrowseComp+~\cite{chen2025BrowseCompPlus}.
We sample 100 queries per dataset: ingesting every query's gold documents makes the full benchmarks too costly to run. 
All sampled queries share one corpus, therefore each query is still resolved against the others' documents as distractors. 
Here we order them by increasing corpus adversariality.

\parai{KILT-HotpotQA} is built from Wikipedia.
Each question is a two-hop \emph{bridge} or a \emph{comparison} question grounded in exactly two gold pages, and natively ships sentence-level provenance. 
Evidence is thus multi-document but the corpus is topically diverse with no near-duplicate passages, isolating cross-document composition from disambiguation.

\parai{BrowseComp+} replaces BrowseComp's live-web
setting with a fixed corpus of $\sim$100K human-verified web documents, labeling per query the \emph{evidence} documents needed to answer and the \emph{gold} documents that also contain the answer. 
These labels are document-level only; therefore we construct character-level evidence spans ourselves to support our attribution analysis (\autoref{app:query-preprocess}).
Its queries are long, obfuscated multi-clue puzzles spanning a broad topical mix and are the most evidence-hungry.
We use a median of 6 (up to 10) evidence documents, stressing the breadth of routing over heterogeneous open-domain text.

\parai{LOONG} is built from financial 10-K filings and spans single-span factual lookups (L1) and multi-span comparative questions (L2) that aggregate a metric (e.g., total operating revenue) across up to nine companies.
Answering an L2 question therefore requires retrieving \emph{every} required evidence span.\footnote{We use 50 L1 and 50 L2 queries: after filtering LOONG to English non-\texttt{paper} L1/L2 questions (\autoref{app:query-preprocess}), only 75 L1 and 80 L2 qualify, and the evidence-clean L2 pool tops out near 50.
Thus, we allocate 50 queries for each L1/L2 type in our testset.} 
Each question bundles several companies' filings.
Unlike the other two datasets, the distractors here are structurally near-identical to the gold document, the same financial line items recur across companies.
This makes telling similar documents apart the central retrieval challenge.
Thus, for passages with near-duplicate table rows, we additionally prepend each chunk with its filing entity (company) name before embedding.

\para{Baselines.} A wrong answer from the full system has two possible causes: retrieval failed to surface the relevant document, or the synthesizer was handed the right document but still answered incorrectly. 
To attribute errors between these two causes, we compare against two baselines that isolate them: one that removes retrieval entirely by always giving the synthesizer perfect evidence, and another that replaces our curated retrieval with an unstructured, realistic alternative.

\parai{Centralized retrieval} represents a common industry practice today (\autoref{sec:current-research}): a single provider crawls, chunks, and indexes content into one shared pipeline, the same architecture used by production systems such as Tavily~\cite{tavily_aws2026}.
We implement it with the same chunking, embedding, and synthesis setting as \proposed, so the two differ only in how retrieval is structured, not in model or preprocessing choices (\autoref{app:centralized-search}).

\parai{Oracle (gold docs given)} isolates retrieval from reading: for each query it skips routing and chunk retrieval and feeds the full gold document(s) directly to the synthesizer, which answers with the same model and synthesis prompt as the full system and is scored with the same correctness metric. 
It therefore measures synthesizer performance when the relevant document is guaranteed to be in context, letting us attribute the full system's errors to retrieval versus reading. 
We report each dataset at its operating retrieval depth $k$: the smallest $k$ at which the system's correctness plateaus (additional chunks no longer improve it).

\parai{Web search agent} is the complementary, realistic baseline: it replaces the curated corpus dataset and routing with an off-the-shelf web-search agent, where the same synthesizer model (\texttt{gpt-5-mini}) answers each query using live public-web search (recording the pages it cites), judged with the same correctness metric. 
It captures what a general LLM-plus-web-search system achieves \emph{without} a purpose-built semantic internet.
We report the agent in a \emph{search-only, no memory} configuration.
It is forced to issue a web search on every query and instructed to answer \emph{only} from the retrieved results, without parametric memory and without deferring via clarification.

As a robustness check, we also evaluate the agent's default \emph{autonomous} behavior, where it decides for itself whether to skip search, answer from memory, or ask for clarification, as a deployed system like ChatGPT or Perplexity.
Correctness differs from the reported search-only figure by at most 4 points on any dataset, so which configuration we report does not change our conclusions (\autoref{app:web-search-agent}).

\para{Implementation Variants.}
We implement resolvers (for both status quo and \proposed) with two design variants.

\begin{smartlist}
    \item \texttt{Base}: semantic search with standard chunking and no contextualization.
    \item \texttt{+Ctx}: \texttt{Base} plus static contextualization (one LLM-generated document summary reused for all chunks from that document).
    Prepending LLM-generated context to chunks before embeddings is now standard practice for improving retrieval, following Anthropic's Contextual Retrieval Technique~\cite{anthropic2024contextual}.\footnote{Our \texttt{+ctx} variant uses one document-level summary rather than a per-chunk summary, trading some retrieval quality for a lower and more predictable ingestion cost.}
\end{smartlist}

We also implement and evaluate a knowledge-graph-based resolver variant; we do not report it as a main-results baseline because it is substantially costlier to ingest and our implementation did not improve correctness (~\autoref{app:kg-search}).

\para{Document Placement Policy.}
Documents are sharded across nodes by a deterministic hash of a per-document \emph{source-identity key}, so that documents sharing an identity colocate on one node and a multi-document query's evidence is spread across nodes, exercising decentralized routing. 
We use a single policy across benchmarks and let the
identity key be whatever stable handle each dataset exposes: the filing entity (company name) for LOONG, the source web domain for BrowseComp+, and the Wikipedia page identity for KILT-HotpotQA.

\begin{figure*}[t]
    \centering
    \includegraphics[width=\textwidth]{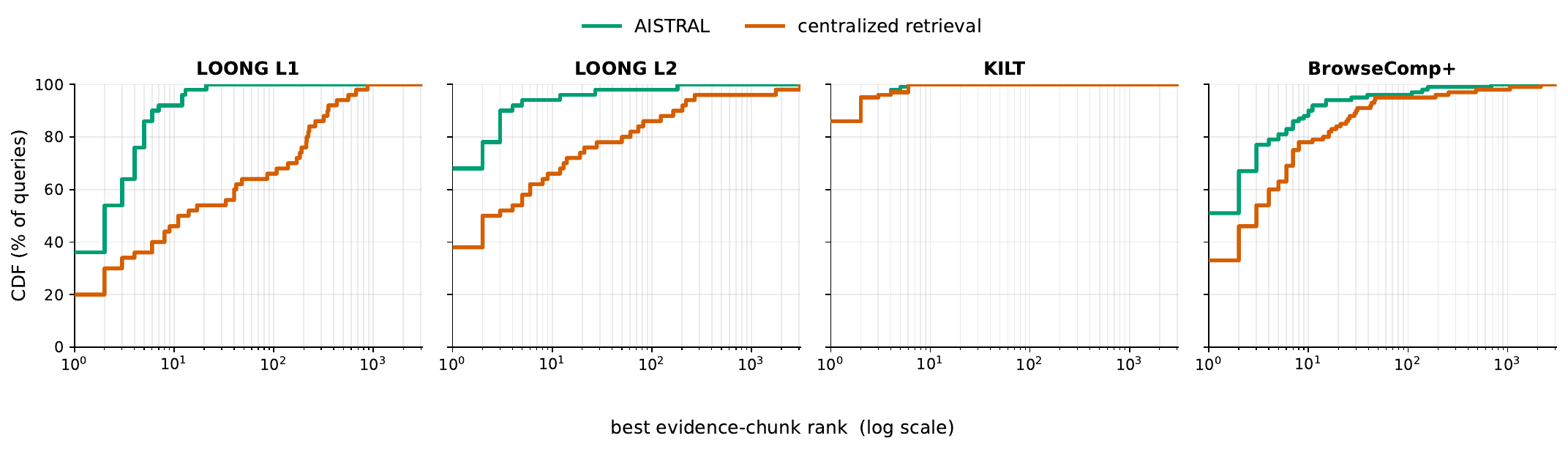}
    \caption{\textbf{Best evidence-chunk rank, per query.} For each query,
    the rank (log scale) of the highest-ranked chunk containing any part of
    the required evidence. 
    Curves show the CDF of this rank across queries; further left is better.
    This measures how quickly the \emph{first} relevant chunk surfaces.}
    \label{fig:rank-cdf}
\end{figure*}

\subsection{Correctness vs.\ Retrieval Traffic}
\label{sec:correctness-vs-traffic}

\para{Deeper retrieval improves correctness, up to a plateau.}
\autoref{fig:tradeoff-k-sweep} (top) shows correctness rising with retrieval depth $k$ and then flattening; we report each dataset at this plateau (shaded), indicating the smallest depth beyond which more chunks no longer help. 
There, \proposed comes close to the perfect-document-retrieval ceiling.
That is, once $k$ is large enough, retrieval is rarely the bottleneck. 

It is worth noting LOONG L2 is an exception: \proposed \emph{exceeds} perfect-retrieval baseline ($0.63$ vs.\ $0.47$), because the full L2 gold context is simply too large for the synthesizer, the synthesis component introduced in~\autoref{sec:system-overview} (\Circled{Q4} in~\autoref{fig:aistral-architecture}).
LOONG's filings are far bigger than the other corpora (median $\sim$150\,KB, vs.\ $\sim$11\,KB Wikipedia pages), and an L2 comparison query draws on a median of three of them at once ($\sim$87\,K tokens, up to $292$\,K).
The enormous context size is enough to overrun the synthesizer's input limit\footnote{\texttt{gpt-5-mini} accepts up to $\sim$272\,K input tokens \cite{openai2025gpt5developers}.} for one query on gold documents alone, leading to context confusion~\cite{liu2024lost, levy2024same}.
In \proposed, chunk retrieval sidesteps this by feeding the synthesizer only the relevant spans.
We omit the perfect-retrieval curve for LOONG L2 from \autoref{fig:tradeoff-k-sweep}: because the gold context itself induces context confusion, it no longer functions as a meaningful upper bound.

\para{\proposed beats both baselines, and the gap grows with corpus difficulty.}
Against a \emph{status quo} retriever, \proposed performs roughly on par on KILT, clearly ahead on BrowseComp+, and dominant on LOONG (three-to-four times more accurate on both levels).
Against a \emph{web-search agent} over the public web, the margin depends on whether the answer is publicly available: on KILT (Wikipedia trivia) the agent nearly matches \proposed, but drops sharply on LOONG and BrowseComp+.
LOONG's drop is largely source disagreement, not retrieval failure, where the public web often has multiple defensible figures for the same metric (\autoref{app:loong-failure-analysis}).
BrowseComp+ is different: it is built so no single public page carries all the clues to answer the query (\autoref{app:bcp-failure-analysis}).

The cause is common to both comparisons: \emph{routing shrinks the pool each query must search}. 
A status quo or public-web retriever ranks the evidence against the whole collection, so on a large, internally repetitive corpus the right chunks are buried among many look-alikes and fall out of the top-$k$.
\autoref{fig:rank-cdf} makes this concrete at the level of individual queries.
For each query, we track the rank of the single best-ranked chunk containing any part of its evidence, and plot its CDF under both routing schemes.
On LOONG and BrowseComp+, status quo's curve trails \proposed's by orders of magnitude in rank, where a sizable fraction of queries see no relevant chunk at all until rank $10^2$--$10^3$, further away from any $k$ in \autoref{fig:tradeoff-k-sweep}.
On KILT the two curves nearly overlap, matching the negligible gap we report there.
This gap is measured under vector-based semantic retrieval specifically, and with stronger retrieval technique, including reranking, hybrid search, or GraphRAG, could narrow it for both \proposed and the status quo alike. 

A larger, unified index doesn't simply transfer to ``more information available''.
Retrieval capacity is a bounded resource that gets consumed by corpus pool size and topical density.
So a single index spanning all documents can be a strictly harder search problem than several smaller, disjoint ones~\cite{reimers2021curse, cuconasu2024power}.
Instead, \proposed confines each query to the twin(s) hosting its documents, so the evidence surfaces easier. 
The effect is small on a small, diverse corpus (KILT) and largest on LOONG's near-templated filings, where every document contributes similar-looking sections.

\para{Contextual chunking tracks base.}
The dashed \proposed{}(+ctx) curves in \autoref{fig:tradeoff-k-sweep} stay close to base on every panel, achieving comparable correctness at a slightly higher traffic cost. 
Both variants clearly outperform the web-search agent everywhere except KILT, where correctness scores saturate quickly across all implementations.
We therefore carry base as the default and revisit what \emph{+ctx} reveals about the metric itself in \autoref{sec:funnel}.

\para{The accuracy costs more traffic, but far less than an open-web agent.}
The correctness boost doesn't come free: retrieval traffic grows with $k$ (\autoref{fig:tradeoff-k-sweep}, bottom), so operating at the plateau rather than the deepest $k$ keeps data movement near its minimum for a given accuracy.
Even so, \proposed requires far less data movement than the web agent.
For example, on LOONG~L2 it transfers only $\sim$0.15\,MB per query at the operating $k$, roughly $40\times$ less than the agent.
The public web is thus both less accurate on specialized corpora and far more bandwidth-hungry; a curated, routed index answers the same queries with a fraction of the data.

\subsection{Where queries fail and where correctness diverges}
\label{sec:funnel}

\begin{figure*}[t!]
    \centering
    \includegraphics[width=0.95\linewidth]{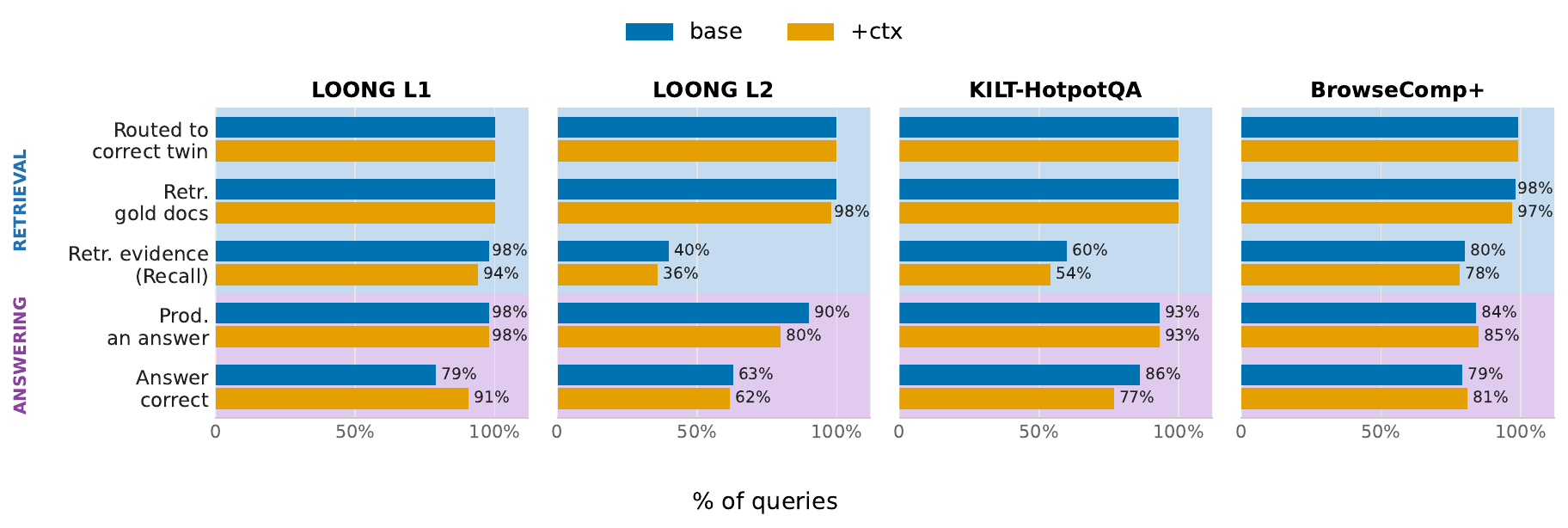}
    \caption{\textbf{Retrieval$\rightarrow$answer attribution for \proposed: base vs.\ contextual chunking.}
    Each stage shows two bars, base with blue and \emph{+ctx} with orange, as the percentage of queries surviving that stage at the plateau-$k$ operating point.
    Stages split into a \emph{retrieval} regime (routing to the gold twin, gold-document retrieval, evidence retrieval; shaded blue), controlled by \proposed, and an \emph{answering} regime (producing an answer, answering correctly; shaded purple), controlled by the synthesizer LLM.
    Routing and document retrieval are near-saturated on all benchmarks; the dominant loss is \textbf{recall}.
    Contextual chunking lowers it yet does not lower correctness in step --- the retrieval and answer stages move independently.}
    \label{fig:fine-grained-funnel}
\end{figure*}

\autoref{sec:correctness-vs-traffic} showed \proposed overall accuracy and cost; this section asks \emph{why}.
We break each query into its retrieval and answering stages to find where correctness is lost, and check whether high correctness always means retrieval worked.

\autoref{fig:fine-grained-funnel} traces each query through the pipeline, stage by stage.
Routing and gold-document retrieval saturate (${\sim}100\%$ on every benchmark), so neither is the bottleneck\footnote{We do not evaluate how routing accuracy degrades as the number of twins grows, for budget and time concerns.
\autoref{sec:discussion} discusses this as an open scaling question.}.
The loss concentrates at one stage: \textbf{recall}, the fraction of required evidence retrieved (\autoref{app:query-preprocess}). 

It falls to $40$--$60\%$ on the LOONG L2 and KILT, and to 80\% on BrowseComp+.
Getting the right document is easy; getting every needed passage into the top-$k$ is where retrieval falls short\footnote{Not specific to our method: replacing dense embeddings with BM25 or a knowledge-graph retriever (\autoref{app:kg-search}) performed worse, not better. 
BM25 missed $\approx20\%$ more required passages, and the KG retriever answered LOONG questions correctly $\approx 18\times$ less often. 
Decomposition, entity-name prefixing, and twin-scoped search help instead.}.

\para{Correctness diverges from recall, but behaves differently across datasets.}
Final correctness is shaped by two separable factors: how effectively the pipeline retrieves the required evidence, and how much the synthesizer can independently answer from its own parametric knowledge, regardless of what retrieval supplies.
To isolate the second factor, we run a closed-book ablation: withholding all retrieved context and scoring what the synthesizer answers from parametric knowledge alone (implementation and full breakdown in \autoref{app:closed-book}). 
This lets us categorize each dataset by how it can be answered short of complete, correct retrieval.

\begin{smartlist}
    \item \textbf{Retrieval-required} (BrowseComp+): closed-book correctness is near zero.
    This means the answer cannot be guessed or recalled without retrieval. 
    Correctness is retrieval-bound.
    \item \textbf{Guessable-by-elimination} (LOONG L2): 
    closed-book correctness is also near zero, so the synthesizer cannot be recalling the figures from memory.
    But the question itself narrows the answer space: a comparison of the form ``among A, B, C, who had the highest X?'' only has as many possible answers as candidates named in the question. 
    Even with incomplete retrieval it often lands on the right one: in 70\% of these cases the winning company's name already appears in a retrieved passage, so it reads the answer off rather than computing it (the same single-hop artifact as KILT); in the remaining 30\% it has neither the values nor the name and simply guesses correctly among the few candidates. 
    Correctness thus runs ahead of recall with no memory involved. 
    \item \textbf{Answerable-from-memory} (KILT): closed-book correctness is high (67\%), since HotpotQA is sourced from Wikipedia and much of it is answerable from parametric knowledge alone. 
    Pipeline correctness ($0.86$) is only $\sim$19 points above this ceiling, so retrieval is adding comparatively little. 
    Unlike LOONG L2's candidate-constrained guessing, KILT's multi-hop questions can be solved by two additional shortcuts: a well-documented single-hop artifact, where the answer text appears in one retrieved passage even though not every annotated bridge hop was fetched~\cite{min2019compositional, jiang2019avoiding} (76\%); and, more directly, cases where the \emph{final} hop is never retrieved at all and the synthesizer supplies it from memory (24\%).
\end{smartlist}

LOONG L1 falls outside this categorization: retrieval is already saturated, so there is no divergence to explain: the gap to correctness (98\% $\rightarrow$ 79\%) is entirely synthesis error.

\section{Discussion}
\label{sec:discussion}
\para{Governance of the Semantic Resolver:}
Who operates and governs the semantic resolver remains an open question. Unlike DNS, semantic resolution raises new challenges around trust, transparency, and ranking bias, and whether it is run by a consortium, federated providers, or a decentralized network will fundamentally affect neutrality and adoption.
Viable designs would require auditable response logs, transparent ranking signals, and open evaluation frameworks to detect bias, abuse, or capture.
Industry is already heading in this direction: Cloudflare's AI Index pairs per-publisher indexes with an aggregated cross-site \emph{Open Index}~\cite{cloudflare_aiindex2025}, and \emph{Pay Per Crawl}~\cite{cloudflare_paypercrawl2025} mechanizes publisher compensation, validating the incentive-alignment premise behind \textbf{O1}.
These, however, remain single-vendor layers rather than an open standard, and the Open Index both discovers and serves content, whereas \proposed's resolver is content-free by contract, with retrieval terminating at the publisher.

\para{Hybrid Centralized--Decentralized Retrieval:}
A separate drawback of \proposed's design is latency: multi-stage retrieval (resolve, then fetch from each source) is inherently slower than querying one centralized index.
A hybrid design can trade some decentralization for speed, indexing recent or frequently-queried content centrally as a cache while the rest stays at the Agentic Twins, keeping publisher control as the default while meeting latency demands for hot content.

\para{Where Trust Enters the Pipeline: Resolver vs. Agent:}
Semantic retrieval naturally splits into two stages, and trust interacts with both.
The resolver returns semantically relevant sources, and may attach provenance metadata and basic trust indicators (e.g., who operates the site, source reputation, and abuse or spam signals~\cite{garzon2025ai}).
AI agents then query the Agentic Twins for fine-grained chunks and may re-rank across sources and chunks using recency, safety filters, and application-specific policies.
Separating source-level retrieval from client-side re-ranking exposes a diverse candidate pool rather than a single opaque ranking, enabling independent auditing and alternative trust strategies while preserving application control over final ordering.

\para{From Advisory Metadata to Enforceable Provenance:}
Publisher-supplied attribution metadata (source, authorship, timestamp) already travels end-to-end, from \texttt{insert()} into the resolver's store and back to the agent at \texttt{search} time, but only as advisory, uninterpreted fields.
Nothing is cryptographically signed or verified, and the copyright and usage-policy fields are not yet populated or enforced.
Making these signals trustworthy is future work: semantic chunks must carry verifiable provenance, including lineage, signatures, and machine-readable attribution, and publishers need standardized ways to express usage constraints that resolvers can enforce. Balancing flexible developer access with strong protection of intellectual property at Internet scale remains an open challenge.

\para{Correctness Is Not Binary on the Public Web:}
Our correctness metric assumes each query has a single gold answer. That holds by construction in a curated corpus, but not on the public web. Our failure analysis (\autoref{app:loong-failure-analysis}) shows the web-search agent often finds the right entity but reports a different, defensible value, e.g., a GAAP figure instead of an adjusted one, or an original filing instead of a later restatement. Each is ``correct'' relative to a different authoritative source, so the metric records a failure where there is really an ambiguity. \proposed narrows this problem rather than solving it. Retrieving only from publisher-registered sources limits \emph{which} documents can conflict, and provenance tells the agent \emph{where} a number came from, a precondition for judging trust. But deciding which of two authoritative numbers is right lies beyond what routing or
registration alone can guarantee.

\para{Publisher Incentives and Operator Buy-in:}
Adoption asks website operators to restructure their content, so they need clear incentives. Some are direct: lower bandwidth from serving compact semantic units instead of full pages, better visibility and attribution in AI-mediated interfaces, richer analytics on how agents
consume content, and the ability to charge AI providers for access through the Twins they operate, offsetting zero-click losses. These incentives are only as sound as the mechanisms beneath them: if the resolver deduplicates near-identical content from distinct publishers, the collapsed publisher loses credit and attribution despite being a legitimate source, so deduplication must be incentive-compatible, not an afterthought. Adoption also needs tooling that lets publishers convert existing sites into AI-native representations, control chunk boundaries and metadata, and attach provenance and policy signals without a full redesign. Ultimately, we believe producing content with AI consumption in mind from the outset will become essential in the age of AI-based search.

\para{Scaling to the Web:}
Our implementation uses a single semantic resolver over 10 Agentic Twin servers. Internet scale raises open problems at both the resolver and the routing layer. No single resolver instance can index the entire web, so the resolver itself must be sharded hierarchically like DNS, from top-level knowledge domains down to specialized sub-domains. At the routing layer, orders of magnitude more Twins increase the chance that a query's evidence fragments across servers, and which Twin should host which document is itself unsolved: our placement policy (\autoref{sec:eval}) is a testbed hash, not a deployment policy. Neither problem is exotic. Partitioning, placement, and
delegated resolution are classic distributed-systems problems, and DNS solved their naming analogues at Internet scale. What is genuinely open is whether those solutions survive the shift from exact-match to semantic resolution, where a query has no unique home and shard
boundaries are soft.

\para{Coexistence with the Existing Web:}
The architecture must integrate with current HTTP infrastructure, CDNs, and publishing workflows, letting publishers expose semantic chunks alongside existing HTML without disrupting legacy traffic.
In our vision, HTML pages remain the primary representation for human browsing while semantic manifests and chunk APIs provide a parallel view for LLMs and agents; how these dual representations coevolve with traditional web search will shape the long-term impact of an \ainet.

\section{Conclusion}
\label{sec:conclusion}
The web's dominant consumers are no longer the human readers it was built for. We have argued that the resulting dysfunction, crawling
without permission and answers without attribution, is an
architectural problem rather than a retrieval problem, and that the
fix is a move from a crawled web to a published one. \proposed makes
that move concrete. Publishers expose semantic units through Agentic
Twins they control, resolvers answer discovery queries without ever
holding content, and AI agents retrieve only the units they need,
directly from sources. Our prototype shows the design is competitive
as well as principled. It matches or exceeds centralized retrieval,
and answers up to $4.2\times$ more questions correctly than a web-search
agent while moving $2.9$--$41\times$ less data per query. \proposed is a
first effort, deliberately so: a small set of roles and interfaces
and one workable instantiation, not a finished standard. The problems
that remain, resolver governance, trust and ranking,
incentive-compatible deployment, and placement at scale, are larger
than any single group, and they are the familiar problems of any
young discovery layer, whose analogues DNS faced at a similar age.
The Semantic Web supplied the ambition a quarter-century ago, and
LLMs have now supplied the consumer. We offer \proposed as a concrete
starting point, and invite the community to refine its interfaces,
contest its design choices, and build the publisher-controlled layer
the web now needs.

\newpage
\bibliographystyle{plainurl}
\bibliography{bibs}
\clearpage
\appendix
\onecolumn
\section{Implementation of Centralized AI-based Websearch}
\label{appen:centralized}
In this paper we consider two different implementations of the centralized AI-based web search. 
A semantic search based implementation and a Knowledge Graph based implementation. 

\subsection{Semantic Search based Centralized AI web search}
\label{app:centralized-search}
In a semantic search based design, all web data resides in a single semantic search entity that maintains a vectorized version of the entire web.
The user's query is sent to this web-scale vector database, which returns the most relevant $top_k$ pieces of information. 

There are two main functions of a semantic resolver: ingestion and query resolution. Our implementation of the processes for ingestion and query resolution are shown in \autoref{algo:centralized-ingest} and \autoref{algo:centralized-query}, respectively.

\paragraph{Ingestion Flow}
The ingestion works by chunking, embedding, and then storing the data from the web servers in a unified vectorDB instance. 

\paragraph{Query Resolution}
When a user submits a query, the semantic resolver performs semantic search using the vectorDB. The relevant filters are applied while performing the search and the $top_k$ relevant chunks are retrieved and returned to the LLM/user in response. Query resolution process can also take a request to return more context around the return chunks. This can be quite useful for more complex queries. 

\begin{algorithm}
\caption{Centralized Semantic Ingestion}
\begin{algorithmic}[1]
\State \textbf{Input:} Document(s) $D$
\State $C \gets \text{Split } D \text{ into chunks}$
\For{each chunk $c_i \in C$}
    \State $m_i \gets \text{ExtractMetadata}(D(c_i))$ \Comment{doc\_id, server\_url, etc.}
    \State $\text{VectorDB.insert}(c_i, m_i)$
\EndFor
\end{algorithmic}
\label{algo:centralized-ingest}
\end{algorithm}

\begin{algorithm}
\caption{Centralized Semantic Resolution}
\begin{algorithmic}[1]
\State \textbf{Input:} Query $q$, expanded\_context $n$
\State $F \gets \text{BuildMetadataFilter}(\text{request})$
\State $C_{top} \gets \text{VectorDB.query}(q, \text{where}=F, \text{limit}=k)$
\State $C_{final} \gets \emptyset$
\For{each $c_i \in C_{top}$}
    \If{$n > 0$}
        \State $S \gets \text{VectorDB.getNeighbors}(c_i.\textit{doc\_id}, c_i.\textit{index} \pm n)$
        \State $C_{final} \gets C_{final} \cup \{\text{Concatenate}(S)\}$
    \Else
        \State $C_{final} \gets C_{final} \cup \{c_i\}$
    \EndIf
\EndFor
\State \textbf{Return} $C_{final}$
\end{algorithmic}
\label{algo:centralized-query}
\end{algorithm}

\subsection{KG based Centralized AI web search}
\label{app:kg-search}
The KG-based semantic resolver combines graph traversal with vector search to retrieve the most relevant pieces of information. 

The process for ingestion and query resolution are shown in \autoref{algo:centralized-graph-ingest} and \autoref{algo:centralized-graph-query}, respectively.

\paragraph{Ingestion Flow}
The ingestion process works by first extracting the graph of entities and relations per document. 
This graph is then merged into a Global Knowledge Graph. 
Each node in the graph stores associated \texttt{doc\_ids} for all the documents that the entity is present in. 
Each document is then chunked, the chunks are converted into embeddings and are stored in a vectorDB along with the metadata for each chunk that also includes the \texttt{doc\_id} that the chunk belongs to. 

\paragraph{Query Resolution}
During the query resolution process, entities are extracted from the query. All the related nodes in the global knowledge graph are extracted via $n$-hop traversal. The \texttt{doc\_ids} in the nodes are then used to create a filter for vector search that retrieves semantically similar chunks. 

\begin{algorithm}
\caption{Centralized KG based Ingestion}
\begin{algorithmic}[1]
\State \textbf{Input:} Documents $D$
\State $G_{global} \gets \emptyset$
\For{each document $d_j \in D$}
    \State $(E_j, R_j) \gets \text{LLM-Extract}(d_j)$ \Comment{Extract graph per document}
    \For{each node $e \in E_j$}
        \State $e.\textit{doc\_ids} \gets e.\textit{doc\_ids} \cup \{\text{id}(d_j)\}$ \Comment{Update document lineage}
    \EndFor
    \State $G_{global} \gets \text{Merge}(G_{global}, E_j, R_j)$ \Comment{Merge into Global Knowledge Graph}
    \State $C_j \gets \text{Split } d_j \text{ into chunks}$
    \For{each chunk $c_{j,k} \in C_j$}
        \State $v_{j,k} \gets \text{Embedding}(c_{j,k})$
        \State $\text{VectorDB.insert}(v_{j,k}, \text{metadata}=\{\text{text}: c_{j,k}, \text{doc\_id}: \text{id}(d_j)\})$
    \EndFor
\EndFor
\end{algorithmic}
\label{algo:centralized-graph-ingest}
\end{algorithm}

\begin{algorithm}
\caption{Centralized KG based query resolution}
\begin{algorithmic}[1]
\State \textbf{Input:} User Query $q$, Global Knowledge Graph $G_{global}$, Hop count $n$
\State $E_q \gets \text{LLM-ExtractEntities}(q)$ 
\State $V_{rel} \gets \text{Traverse}(G_{global}, E_q, n)$ 
\State $D_{filter} \gets \bigcup_{v \in V_{rel}} v.\textit{doc\_ids}$ 
\State $v_q \gets \text{Embedding}(q)$
\State $R \gets \text{VectorDB.search}(v_q, \text{filter}=\{\text{doc\_id} \in D_{filter}\})$ 
\State \textbf{Return} Retrieved chunks $R$
\end{algorithmic}
\label{algo:centralized-graph-query}
\end{algorithm}

\section{Interface Specification}
\label{appen:endpoint-datamodels}

\proposed separates publication, discovery, and retrieval into distinct interfaces so that publishers, semantic resolvers, and AI agents each retain a clear responsibility boundary. 
At a high level, publishers expose content and metadata for semantic preparation, semantic resolvers support source discovery over source-level representations, and agentic web servers support fine-grained retrieval over local semantic units. 

\subsection*{\texttt{publish(content, metadata)}}
The \texttt{publish} interface captures the publisher-facing side of the architecture. 
Its purpose is to hand content and basic source metadata to an agentic web server, which can then prepare local semantic representations for retrieval. 
Conceptually, this is where publisher control enters the system: content is not first crawled and reformatted by a centralized search provider, but rather provided together with metadata chosen by the publisher or content operator.

For the purposes of the design, the most important fields are shown in ~\autoref{tab:publish-schema}. 
These fields express the source identity, the content itself, and the minimal metadata needed for provenance, freshness, and modality-aware handling. 
Additional fields such as licensing or access policy are natural extensions, but are not essential to the core interface decomposition described in this paper.

\begin{table}[h]
\centering
\renewcommand{\arraystretch}{1.18}
\begin{tabular}{p{1.7cm} p{2.3cm} p{2.1cm} p{1.8cm} p{5.4cm}}
\toprule
\textbf{Interface} & \textbf{Argument} & \textbf{Field} & \textbf{Type} & \textbf{Design role} \\
\midrule
\multirow{5}{*}{\texttt{publish}} 
& \texttt{content} & --- & string/object & Content to be semantically prepared and served locally. \\
\cline{2-5}
& \multirow{4}{*}{\texttt{metadata}}
& \texttt{source} & string & Canonical source reference, e.g., URL or document identifier. \\
& & \texttt{title} & string & Human-readable attribution label. \\
& & \texttt{timestamp} & time & Freshness signal for downstream discovery and retrieval. \\
& & \texttt{content\_type} & string & Data modality or representation type. \\
\bottomrule
\end{tabular}
\caption{Design-relevant fields for \texttt{publish(content, metadata)}.}
\label{tab:publish-schema}
\end{table}

\subsection*{\texttt{insert(documents)}}
The \texttt{insert} interface registers content records with the semantic resolver. 
Its role is to make content discoverable without requiring the resolver to serve the content itself. 
In this sense, \texttt{insert} is a source-registration interface rather than a user-facing retrieval interface. 
The argument name \texttt{documents} is retained for readability, but what matters architecturally is that each item carries enough semantic and metadata information to support source discovery and filtering, as listed in \autoref{tab:insert-schema}.

\begin{table}[h]
\centering
\renewcommand{\arraystretch}{1.18}
\begin{tabular}{p{1.7cm} p{2.3cm} p{2.1cm} p{1.8cm} p{5.4cm}}
\toprule
\textbf{Interface} & \textbf{Argument} & \textbf{Field} & \textbf{Type} & \textbf{Design role} \\
\midrule
\multirow{7}{*}{\texttt{insert}}
& \multirow{7}{*}{\texttt{documents}}
& \texttt{content} & string & Semantic representation registered for source discovery. \\
& & \texttt{server\_url} & string & Endpoint of the agentic web server that can later serve retrieval results. \\
& & \texttt{source} & string & Canonical source reference preserved for provenance and attribution. \\
& & \texttt{title} & string & Source or document reference exposed to downstream consumers. \\
& & \texttt{timestamp} & time & Supports freshness-aware discovery constraints. \\
& & \texttt{category} & string & Supports topical partitioning or coarse semantic filtering. \\
& & \texttt{language} & string & Supports language-aware discovery. \\
\bottomrule
\end{tabular}
\caption{Design-relevant fields for \texttt{insert(documents)}.}
\label{tab:insert-schema}
\end{table}

\paragraph{Behavior.}
\proposed supports two resolver realizations that differ in how registered sources are indexed: a \emph{vector-based} resolver and a \emph{knowledge-graph (KG) based} resolver. 
The two share the same \texttt{insert} interface but populate different indices at ingestion time; their query-time behavior is described under \texttt{resolve}.

\emph{Vector-based ingestion.} A crawler crawls the agentic web servers and generates a summary (e.g., using an LLM) of each crawled document or file. 
The summary is vectorized using an embedding model and inserted into a vector database, along with metadata about the agentic twin that holds the chunks of the corresponding document. 
The resolver therefore stores only source-level summaries rather than the underlying content, keeping it a lightweight discovery layer (\autoref{algo:p-decentralized-ingest}).

\emph{Context-enriched (+ctx) ingestion.} The \texttt{+ctx} variant refines the vector-based path with static, document-level context on the chunk representations served for cross-document disambiguation.
Each document is summarized once with an LLM, and that summary is prepended verbatim to every chunk of the document before the chunk is embedded and stored; the summary is cached per document, so enrichment costs a single LLM call per document regardless of its chunk count.
A more fine-grained alternative would generate context dynamically per chunk, tailoring the prepended summary to each chunk rather than sharing one document-level summary across all of them.
This would likely yield better retrieval accuracy, but it requires a separate LLM invocation for every chunk, each conditioned on the full source document, multiplying ingestion cost and latency; we therefore adopt the static document summary as a deliberate cost/quality compromise and leave dynamic, chunk-tailored context to future work.
This step reuses the document summary $s_j$ already computed in \autoref{algo:p-decentralized-ingest} and requires no change to the query path.

\emph{KG-based ingestion.} An LLM extracts entities and relations per document and creates a mini knowledge graph.
Mini KGs are merged into a global KG, with nodes updated to include the \texttt{server\_url} of the agentic web server that holds the relevant chunks. 
Documents are also summarized and the embeddings of their summaries are inserted into the vector database, so that graph traversal and vector similarity can be combined at query time (\autoref{algo:hierarchical-graph-ingest}).

\begin{algorithm}
\caption{\proposed Semantic Ingestion}
\begin{algorithmic}[1]
\State \textbf{Input:} Documents $D$ with source URLs $U$
\For{each $(d_j, u_j) \in (D, U)$}
    \State $m_j \gets \text{ExtractMetadata}(d_j, u_j)$
    \State $s_j \gets \text{LLMSummary}(d_j)$
    \State $\text{VectorDB.insert}(s\_j, m\_j)$
\EndFor
\end{algorithmic}
\label{algo:p-decentralized-ingest}
\end{algorithm}

\begin{algorithm}
\caption{\proposed KG Ingestion}
\begin{algorithmic}[1]
\State \textbf{Input:} Documents $D$ with source URLs $U$
\For{each $(d_j, u_j) \in (D, U)$}
    \State $G_j \gets \text{LLM-ExtractGraph}(d_j)$
    \For{each node $e \in G_j$}
        \State $e.\textit{server\_urls} \gets e.\textit{server\_urls} \cup \{u_j\}$
    \EndFor
    \State $\text{VectorDB.insert}(d_j)$
\EndFor
\State $G_{global} \gets \text{Merge}(G_{global}, \bigcup G_j)$
\end{algorithmic}
\label{algo:hierarchical-graph-ingest}
\end{algorithm}

\subsection*{\texttt{resolve(query, filters)}}
The \texttt{resolve} interface is the discovery or control-plane interface. 
Its purpose is not to return answer content, but to identify the sources that should be queried next.

As shown in \autoref{tab:resolve-schema}, queries in \proposed are accompanied by optional constraints that let the client narrow source discovery by domain, topic, language, or freshness. 

\begin{table}[h]
\centering
\renewcommand{\arraystretch}{1.18}
\begin{tabular}{p{1.7cm} p{1.6cm} p{2.8cm} p{1.8cm} p{5.4cm}}
\toprule
\textbf{Interface} & \textbf{Argument / Return} & \textbf{Field} & \textbf{Type} & \textbf{Design role} \\
\midrule
\multirow{7}{*}{\texttt{resolve}}
& \texttt{query} & --- & string & Query to be mapped to relevant sources. \\
\cline{2-5}
& \multirow{5}{*}{\texttt{filters}}
& \texttt{limit} & integer & Upper bound on the number of returned agentic web server. \\
& & \texttt{allowed\_domains} & list[string] & Restricts discovery to selected server domains or endpoints. \\
& & \texttt{allowed\_categories} & list[string] & Restricts discovery to selected semantic categories. \\
& & \texttt{allowed\_languages} & list[string] & Restricts discovery by language. \\
& & \texttt{min/max\_timestamp} & time range & Restricts discovery by freshness range. \\
\cline{2-5}
& \texttt{return} & \texttt{urls} & list[string] & Agentic web server(s) to query next through \texttt{search(query, limit)}. \\
\bottomrule
\end{tabular}
\caption{Design-relevant fields for \texttt{resolve(query, filters)}.}
\label{tab:resolve-schema}
\end{table}

\paragraph{Behavior.}
\texttt{resolve} takes a query and returns the addresses of the agentic web sources holding semantically relevant data, applying the metadata-based filters above (data format, copyright, timeliness, quality, etc.) when selecting sources. 
Its behavior mirrors the two ingestion variants introduced under \texttt{insert}.

\emph{Vector-based resolution.} The resolver holds a global index over the vectorized summaries of the documents it knows about. 
A query is matched against this index via document-level semantic similarity, and the resulting set of relevant source URLs is returned to the AI agent, which then queries those agentic web servers directly through \texttt{search} and synthesizes an answer from the retrieved chunks (\autoref{algo:p-decentralized-query}).
The \texttt{+ctx} variant leaves this path unchanged: it issues the same query against the same index.
The only difference is that the chunks later retrieved through \texttt{search} carry their document's prepended summary.

\emph{KG-based resolution.} Vector similarity alone does not capture relationships between entities within or across documents and may miss information that depends on relational structure rather than surface similarity. 
Extending the resolver with a KG layer that encodes entities and their relations enables more structured, multi-hop resolution, and provides a natural mechanism for embedding constraints into discovery, such as restricting retrieval to specific entities, relations, sources, time ranges, or publisher-defined policies. 
At query time, entities are extracted from the query using an LLM and expanded via $n$-hop traversal; server URLs recovered from the node metadata are combined with the $top_{k}$ URLs from vector-based search over the document summaries, and the union is returned (\autoref{algo:hierarchical-graph-query}).

We implement this KG layer but report the vector-based resolver as our system: on our benchmarks it did not improve retrieval accuracy over vector similarity, and was substantially worse on LOONG. 
The failure is structural:
\begin{itemize}
  \item \textbf{The answer is not a node.} Our queries resolve to specific quantitative facts stated at single points within long narrative filings (${\sim}66\%$ of LOONG gold answers are numeric values).
  LLM entity--relation extraction captures the narrative's entities but not these numeric statements, so the graph does not hold the evidence the query needs.
  \item \textbf{Merges collide on repetitive corpora.} Across many structurally similar filings, extracted entities and relations overlap, so merging the mini-KGs fuses or fragments nodes and traversal cannot discriminate which source to return.
  \item \textbf{Queries link unreliably to nodes.} A natural-language query rarely matches the surface form of an extracted node, so traversal misses.
  Vector similarity is robust to this paraphrase gap.
  \item \textbf{It costs more for less.} Per-document extraction and global merge are expensive at ingest, and graph traversal is slower at query time.
\end{itemize}
The KG layer's value in our setting is therefore its support for \emph{structured constraints} (entity, relation, source, time, and policy filters), not retrieval accuracy.

\begin{algorithm}
\caption{\proposed Semantic Query Resolution}
\begin{algorithmic}[1]
\State \textbf{Input:} Query $q$
\State $F \gets \text{BuildMetadataFilter}(\text{request})$
\State $R \gets \text{VectorDB.query}(q, \text{where}=F, \text{limit}=k)$
\State $U \gets \{m.\textit{server\_url} \mid m \in \text{Metadatas}(R)\}$
\State \textbf{Return} $\text{Deduplicate}(U)$
\end{algorithmic}
\label{algo:p-decentralized-query}
\end{algorithm}

\begin{algorithm}
\caption{\proposed KG Query Resolution}
\begin{algorithmic}[1]
\State \textbf{Input:} Query $q$
\State $E_q \gets \text{LLM-ExtractEntities}(q)$
\State $E_{expanded} \gets \text{Traverse}(G_{global}, E_q, n)$
\State $U_{graph} \gets \bigcup_{e \in E_{expanded}} e.\textit{server\_urls}$
\State $U_{vector} \gets \text{VectorDB.searchUrls}(q)$
\State \textbf{Return} $\text{Deduplicate}(U_{graph} \cup U_{vector})$
\end{algorithmic}
\label{algo:hierarchical-graph-query}
\end{algorithm}

\subsection*{\texttt{search(query, limit)}}
The \texttt{search} interface is the retrieval or data-plane interface exposed by an agentic web server. 
After a resolver identifies relevant sources, the AI agent uses this interface to retrieve semantically relevant units from a selected source. 
Unlike \texttt{resolve}, this interface returns content-bearing objects rather than source references.

\autoref{tab:search-schema} shows the fields that matter most for understanding this design.
The interface may also return auxiliary local metadata for provenance, interpretability, and implementation-level inspection.

\begin{table}[h]
\centering
\renewcommand{\arraystretch}{1.18}
\begin{tabular}{p{1.5cm} p{1.6cm} p{3cm} p{1.8cm} p{5.4cm}}
\toprule
\textbf{Interface} & \textbf{Argument / Return} & \textbf{Field} & \textbf{Type} & \textbf{Design role} \\
\midrule
\multirow{6}{*}{\texttt{search}}
& \texttt{query} & --- & string & Query to be evaluated within a selected source. \\
\cline{2-5}
& \texttt{limit} & --- & integer & Number of local semantic units to return. \\
\cline{2-5}
& \multirow{4}{*}{\texttt{return}}
& \texttt{document} & string/object & Retrieved semantic unit returned to the AI agent. \\
& & \texttt{metadata} & object & Optional provenance and within-source location metadata for the unit. \\
& & \texttt{score}/\texttt{distance} & float & Optional local ranking signal from lexical or vector retrieval. \\
& & \texttt{before/after\_context} & string & Optional neighboring context for interpretability or expansion. \\
\bottomrule
\end{tabular}
\caption{Design-relevant fields for \texttt{search(query, limit)}.}
\label{tab:search-schema}
\end{table}

\paragraph{Behavior.}
Through \texttt{search}, agentic web servers act as queryable data sources for AI agents: in addition to serving webpages, a server receives a query and replies with the most relevant $top_k$ semantic units, and makes all types of information searchable, including images, documents, audio, video, and databases. 
Because data providers understand their own data better than a centralized system, they retain full control over how content is chunked, embedded, and returned in response to a query, which enables more relevant results while preserving complete data sovereignty. 
This control also opens new business models: data providers can monetize their content by charging commercial AI agents for access while potentially offering free access to human users.

\subsection{AI-Agent Internal Interfaces.}
In addition to the cross-component interfaces above, the AI agent performs two internal operations that are architecturally important for end-to-end behavior: query processing before retrieval and answer synthesis after retrieval. 
These are not public network APIs, but they materially affect how user intent is translated into retrieval actions and how retrieved evidence is converted into a final answer.

\subsection*{\texttt{process\_query(q)}}
The \texttt{process\_query} interface transforms a user query into a retrieval-oriented representation. 
Its role is to surface implicit constraints, reduce ambiguity, and produce query forms that are easier for downstream discovery and retrieval components to use effectively. 
Depending on the task, this stage may expose comparison scope, identify implied constraints, decompose a complex request into multiple subqueries, rephrase the request into a more retrieval-friendly form, or anonymize sensitive information before retrieval. 
In the evaluation setting, this interface is especially important for under-specified queries, such as comparison questions among a fixed range of companies statistics in LOONG benchmark.

\begin{table}[h]
\centering
\renewcommand{\arraystretch}{1.18}
\begin{tabular}{p{1.9cm} p{1.6cm} p{2.6cm} p{1.8cm} p{5.4cm}}
\toprule
\textbf{Interface} & \textbf{Argument / Return} & \textbf{Field} & \textbf{Type} & \textbf{Design role} \\
\midrule
\multirow{5}{*}{\texttt{process\_query}}
& \texttt{query} & --- & string & Original user query. \\
\cline{2-5}
& \multirow{4}{*}{\texttt{return}}
& \texttt{sanitized\_query} & string & Rewritten query used for downstream discovery and retrieval. \\
& & \texttt{subqueries} & list[string] & Decomposes a complex request into simpler retrieval steps. \\
& & \texttt{constraints} & object & Surfaces implicit requirements such as scope, freshness, or grounding needs. \\
\bottomrule
\end{tabular}
\caption{Design-relevant view of \texttt{process\_query(query)}.}
\label{tab:process-query}
\end{table}

\paragraph{Behavior.}
While a user query can in principle be used directly for discovery, processing it first improves efficiency and effectiveness and can reduce exposure of Personally Identifiable Information (PII). 
Processing comprises one or more of: 
\begin{smartlist}
\item \emph{decomposition} of a complex query into simpler subqueries --- possibly iteratively, so that an answered subquery can further decompose or disambiguate the remainder;
\item \emph{anonymization} to remove PII; and
\item \emph{rephrasing} to remove ambiguity and optimize for search.
\end{smartlist}
Processing also extracts the functional requirements implicit in free-form text and turns them into filters for downstream discovery. 
For example, \emph{intent classification} narrows the agentic web sources worth searching.
A factoid lookup needs few retrievals, whereas a qualitative analysis needs deeper research.
A query can also target a specific modality such as video or image.
Likewise, a \emph{freshness} requirement extracted from the query maps onto the timestamp filters exposed by \texttt{resolve}.

\subsection*{\texttt{synthesize(query, context)}}
The \texttt{synthesize} interface converts retrieved evidence into a final answer. 
Given a query and retrieved context, \texttt{synthesize} combines relevant semantic units, resolves partial or conflicting evidence, and produces the final grounded answer.

\begin{table}[h]
\centering
\renewcommand{\arraystretch}{1.18}
\begin{tabular}{p{1.9cm} p{1.6cm} p{2.4cm} p{1.8cm} p{5.4cm}}
\toprule
\textbf{Interface} & \textbf{Argument / Return} & \textbf{Field} & \textbf{Type} & \textbf{Design role} \\
\midrule
\multirow{5}{*}{\texttt{synthesize}}
& \texttt{query} & --- & string & Original or processed query whose answer is to be produced. \\
\cline{2-5}
& \texttt{context} & --- & list/object & Retrieved semantic units used for answer construction. \\
\cline{2-5}
& \multirow{3}{*}{\texttt{return}}
& \texttt{answer} & string & Final answer derived from the retrieved context. \\
& & \texttt{citations} & list/object & Optional evidence references preserved for grounding or attribution. \\
\bottomrule
\end{tabular}
\caption{Design-relevant view of \texttt{synthesize(query, context)}.}
\label{tab:synthesize}
\end{table}

\subsection*{Extensions and Open API Work}
The interfaces above capture the core architectural decomposition of \proposed, but they do not yet constitute a complete production protocol suite. 
Several extensions remain important future work, including explicit \texttt{update} and \texttt{deletion} semantics, richer policy and licensing metadata, stronger provenance and capability descriptors in \texttt{resolve}, access control and authorization, pagination or continuation for long retrieval results, and standardized error schemas. 
These additions are important for deployment and interoperability, but are orthogonal to the main architectural point of the design: separating publication, registration, discovery, and fine-grained retrieval into explicit interfaces with distinct responsibilities.

\paragraph{Semantic caching.}
Any Internet-scale deployment of \proposed will require caching. 
Because the architecture exposes distinct interfaces, caching is naturally \emph{layered}: it can be applied at \texttt{process\_query} (processed queries and subqueries), at \texttt{resolve} (discovered source sets), and at \texttt{search} (retrieved semantic units). 
Query decomposition and rephrasing further increase hit rates by exposing shared subqueries across complex requests. 
Prior work such as Cortex~\cite{ruan2025cortex}, Category-Aware Semantic Caching~\cite{wang2025categoryaware}, and domain-specific embedding models~\cite{gill2025advancing} provides a foundation, but a full treatment of cache placement, invalidation, and freshness across these layers is left to future work.

\section{Query Preprocess}
\label{app:query-preprocess}

\begin{table}[t]
\centering
\small
\begin{tabular}{@{}llll@{}}
\toprule
Dataset & Query selection & Query transform & Gold evidence \\
\midrule
KILT-HotpotQA & 50/50 bridge/comparison & none (verbatim) & native (offset rebased) \\
LOONG         & 50/50 L1/L2, evict+replace & entity-set prepend & constructed (staged heuristic+LLM) \\
BrowseComp+ & stratified to domain dist. & none (verbatim) & constructed (heuristic/LLM locator) \\
\bottomrule
\end{tabular}
\caption{Per-dataset preprocessing. Effort scales inversely with native support:
KILT and BrowseComp+ ship usable question formats and evidence; LOONG ships neither.}
\label{tab:preprocess}
\end{table}

We map all three benchmarks onto a single evaluation interface: each query is reduced to a \emph{question}, a set of \emph{gold documents}, and one or more \emph{gold-evidence spans} expressed as file-level character offsets \texttt{(doc, char\_start, char\_end)} into the materialized document text.
The retrieval corpus is the union of the evaluated queries' documents. 
The benchmarks differ in how much of the original datasets supply this evaluation interface natively, so the preprocessing effort differs accordingly (as shown in~\autoref{tab:preprocess}). 
KILT ships evidence natively, whereas LOONG and BrowseComp+ require us to construct it, and LOONG additionally requires rewriting its questions.

\subsection{KILT-HotpotQA}
Selection is deterministic. 
We take the dev split and recover each question's \texttt{bridge}/\texttt{comparison} type by joining on instance id (according to the original HotpotQA release), then seed-sample a balanced
50/50 set. 
Questions are used \emph{verbatim}. 
Each gold Wikipedia page is materialized from the KILT knowledge source where a page's body is stored as a \emph{list} of paragraph strings; we join that list into one contiguous text file (one page $\rightarrow$ one document).

Gold evidence is \emph{native}: KILT provenance gives paragraph-relative offsets \texttt{(paragraph\_id, char)}, which we rebase to file-level offsets by prefix-summing paragraph lengths, $\mathit{char\_start}=\sum_{i<p}\lvert \text{paragraph}_i\rvert + c$. 
A query carries all its provenance spans (2--8 per question). 

\subsection{LOONG}
\paragraph{Filtering.} From the 1{,}600 LOONG questions we keep English rows, drop the \texttt{paper} type, and drop Levels 3--4, leaving 75 Level-1 and 80 Level-2 questions, all over financial 10-K filings.

\paragraph{Query transformation.} Most Level-2 questions are comparative aggregations whose company set is \emph{implicit} --- present only in the filings bundled with the question (``the documents above''). 
To make them answerable under open retrieval, we deterministically rewrite such questions by prepending the candidate entity list, ``Among the following candidate companies: $A$, $B$, \dots'', where the entities are extracted from the bundled document identifiers. 
The rewrite is rule-based and applies only to the underspecified comparative templates.
For the \emph{Practical web search} baseline only, we additionally qualify each LOONG question with its reporting period (fiscal year 2024, recovered from the filing identifiers).

\paragraph{Evidence construction.} LOONG provides no evidence spans, so we build them. 
A natural alternative would be handing the documents to an LLM and having it mark the
evidence directly.
However, this is infeasible at LOONG's scale: the corpus for these queries is several million tokens ($\sim$20$\times$ a 272K-token input window), and even a \emph{single} query's bundled filings (2--9 documents) average ${\sim}136$K tokens and exceed the input context window cap for 12\% of queries.
Moreover, LLMs are unreliable at emitting exact character offsets over long documents due to context confusion. 
We therefore use a staged pipeline in which a heuristic step narrows down the evidence search range.
Stage~1 localizes candidate windows heuristically (metric-term and answer-value matches) within each document; a triage step accepts confident rows.
The remaining rows escalate to Stage~2, a bounded LLM pass (\texttt{gpt-5-mini}) over only the disputed slices; a final deterministic step recovers exact spans by matching the extracted evidence text back into the source document to resolve the dispute. 
Finally, because the bundled filings are structurally near-identical, at ingest we prefix each passage with its source company name so the retriever can disambiguate otherwise-similar rows.

\subsection{BrowseComp+}
\paragraph{Selection.} BrowseComp+ exposes no per-query domain metadata, neither from the original BrowseComp dataset, so we recover a primary domain per query using an eleven-bucket taxonomy interpreted from BrowseComp's published category distribution~\cite{openai2025browsecomp}, labeling by the answer-seeking target rather than incidental clues. 
We then stratified-sample the evaluation slice to match the original BrowseComp topic distribution~\cite{openai2025browsecomp}. 
Questions are used \emph{verbatim}.

\paragraph{Documents and evidence.} The benchmark supplies gold documents and document-level evidence labels; it does \emph{not} supply character-level evidence spans, which we construct with a heuristic locator (numeric and reverse-clue matching) that falls back to an LLM (\texttt{gpt-5-mini}) on hard rows, with handling for answers that recur across a document. 
It is worth noting that BrowseComp+ also ships curated distractor documents; because these cannot be verified as non-contributing, we treat them as optional inputs and exclude them from the evaluated corpus rather than labeling them hard negatives.

\paragraph{Answer-repeat deduplication.} BrowseComp+'s corpus deliberately repeats each answer across multiple documents (median 2 copies per query, up to 10), so retrieving every labeled copy is a stricter requirement than retrieving the answer itself.
\emph{Any} single copy is sufficient to answer the query. 
We therefore score recall as satisfied once \emph{any one} copy is retrieved, rather than requiring all of them. 
This changes the reported recall rate (\autoref{sec:funnel}) from 44\% under the stricter, all-copies criterion to 80\% under the corrected one; end-to-end correctness (79\%) is unaffected, since it depends only on the final answer, not on which copy was retrieved.

\section{Evaluation with Open-ended Queries}
\label{app:open-end-eval}

\begin{figure*}[t!]
    \centering    \includegraphics[width=0.95\linewidth]{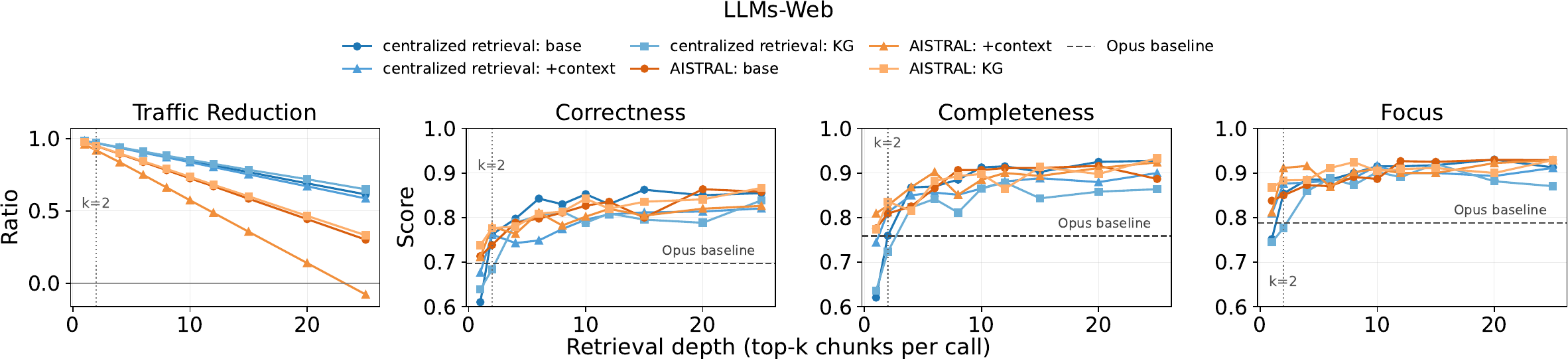}
    \caption{LLMs-and-Web $k$-sweep results under varying retrieval depth. 
    The left panel shows traffic reduction as the per-query retrieval budget $k$ increases, while the three right panels show answer quality in terms of \textit{correctness}, \textit{completeness}, and \textit{focus}. 
    Across both \proposed and centralized variants, most answer-quality gains saturate by small $k$, while larger $k$ primarily reduces traffic efficiency. 
    The dashed horizontal line marks the Opus reference baseline, and the vertical marker highlights $k=2$ crossing the baseline performance.}
    \label{fig:answer-quality-k-sweep}
\end{figure*}

We also evaluate \proposed on LLMs-and-Web~\cite{webconf26}, a benchmark of open-ended natural-language web queries.
It contains a wild variety of queries, ranging from short pointer-style lookups (e.g., locating a site's official page) to informational questions.
Each query is grounded in a set of real web documents and their source URLs harvested from the query's prompt and from web search.
Unlike our closed-ended benchmarks, its queries have no single gold answer, so correctness cannot be scored against a localized evidence span.
We therefore judge each of the 50 evaluated queries with an LLM along three reference-free axes: \emph{correctness} (factual accuracy and plausibility), \emph{completeness} (coverage of what the query asks), and \emph{focus} (staying on-topic and concise).
As a reference point, we run the dataset's gold answers (produced by the model \texttt{Opus}) through the same judge and axes.
We mark the baseline score as a dashed line in~\autoref{fig:answer-quality-k-sweep}, so crossing it means matching the reference answers under the same grader.
This setting complements our main results by testing whether decentralized routing and synthesis hold up when answers are open-ended and graded on quality.

All variants clear the \texttt{Opus} baseline on every axis by $k{\approx}2$ and then flatten.
That is, \proposed saves the most traffic for the same quality.
We report this in the appendix rather than the main results for two reasons.
First, the open-ended scores are hard to read as retrieval signal: the axes saturate near their capability once a few relevant chunks are fetched. 
There the differences between variants are dominated by judge noise, with no principled meaning to attach to, say, a completeness of $0.90$ versus $0.85$.

Second, and more importantly, this benchmark has no localized gold evidence, so we cannot run the stage-by-stage attribution of~\autoref{sec:funnel}: when a query scores low we cannot tell whether routing, span retrieval, or synthesis is at fault.
As that section shows, the end-to-end score conflates the retriever and the reader; without intermediate signals it is not a debuggable measure of \proposed itself, which is exactly why our main evaluation uses closed-ended benchmarks with span-level gold.

\section{Web Search Agent Baseline}
\label{app:web-search-agent}

\subsection{Configurations}
\label{app:web-search-config}

The web-search agent baseline answers each benchmark query with the same synthesizer model used throughout the paper (\texttt{gpt-5-mini}), equipped with a live web-search tool (the OpenAI Responses \texttt{web\_search} tool).
It receives no benchmark corpus and no routing (only the query and the public web) and is scored with the identical reference-anchored correctness judge applied to every other system.
For each query we record the pages the agent cites and their sizes, giving the per-query traffic figures reported in the main text. 
All runs use the same 100-query subset per benchmark as the retrieval systems.

We evaluate two configurations, differing only in when the agent searches and what it may answer from:
\begin{smartlist}
    \item \textbf{Search-only (no memory):} the configuration reported in the main results.
    The web-search tool is forced on every query (via \texttt{tool\_choice}), and the agent is instructed to answer using only the retrieved results, not to fall back on parametric knowledge, and to commit to a single answer rather than requesting clarification. 
    This mirrors the retrieval systems, which likewise retrieve and then answer every query, isolating the quality of open-web retrieval from the model's latent knowledge.
    \item \textbf{Autonomous:} the agent's default deployed behavior. The tool is offered, but the model decides for itself whether to invoke it.
    It may answer directly from parametric memory, and it may reply with a clarification request instead of an answer. 
    This reflects how a shipped assistant (e.g., ChatGPT or Perplexity) behaves.
\end{smartlist}

\para{Search behavior.} Under the autonomous configuration, the agent produces a substantive answer on 82\% of queries (the remainder are clarification requests or errors); of those answers, 83\% are grounded in at least one cited page and 17\% are given from parametric memory with no citation. 
The memory-only fraction is highly benchmark-dependent: 32\% of answers on KILT-HotpotQA but only 3\% on LOONG.
Tracking the closed-book results of ~\autoref{app:closed-book}: KILT questions are largely answerable from parametric knowledge, whereas LOONG's micro-cap financial figures are not. 
The search-only configuration, by construction, searches on essentially every query (cited-source rate 96--100\%) and never defers.

The exact system/tool prompts for both configurations are given in ~\autoref{app:prompts} and are not reproduced here.

\para{Effect on correctness.} Forcing search and forbidding memory changes correctness within 4 points on any benchmark, as shown in ~\autoref{tab:web-agent-configs}:

\begin{table}[h]
\centering
\begin{tabular}{lccc}
\toprule
Benchmark & Search-only (reported) & Autonomous & $\Delta$ \\
\midrule
LOONG L1       & 0.190 & 0.174 & $+0.016$ \\
LOONG L2       & 0.540 & 0.500 & $+0.040$ \\
KILT-HotpotQA  & 0.857 & 0.830 & $+0.027$ \\
BrowseComp+    & 0.377 & 0.394 & $-0.017$ \\
\bottomrule
\end{tabular}
\caption{Correctness under the search-only (no-memory) configuration vs.\ the agent's default autonomous behavior.}
\label{tab:web-agent-configs}
\end{table}

The two configurations therefore agree to within noise, and our conclusions do not depend on which one is reported.

\subsection{Failure Analysis on LOONG}
\label{app:loong-failure-analysis}

To verify that the low practical-web-search scores on LOONG (\autoref{fig:tradeoff-k-sweep}) reflect a genuine retrieval limitation rather than a question-phrasing artifact, we audit the baseline's failures at both levels.

\para{L1 failure mode (0.17).} On the 38/50 zero-scored queries, the agent reliably identifies the correct company but reports the wrong figure.
We identify three recurring sources of disagreement:
\begin{smartlist}
    \item \emph{Conflicting sources:} the original 10-K, a later 10-K/A restatement, GAAP vs. adjusted figures, and third-party aggregators (e.g., Macrotrends, StockAnalysis, WSJ) each report a different value for the same nominal metric.
    \item \emph{Corporate actions:} reverse stock splits and share-count changes swing per-share figures by orders of magnitude.
    For example, AIM ImmunoTech's Basic EPS is $-\$0.12$ in the gold filing vs. $-\$30.92$ reported on the open web.
    \item \emph{Metric-definition scope:} For example, ``administrative expenses'' is $\$10,135K$ in the gold filing but $\$76,306K$ on the web, summing continuing and discontinued operations.
    ``Capital surplus'' is conflated with the web's ``additional paid-in capital.''
\end{smartlist}

Even with the fiscal year added to each query, the agent cannot choose among several disagreeing public figures.
Only retrieval over the authoritative filing itself resolves this, which underlines the importance of source provenance.

\para{L2 failure mode (0.50).} L2 compounds the L1 problem: a comparison across up to nine companies requires the correct per-company figure for each of them, so any single L1-style error can flip the comparison.
L2 nonetheless scores higher than L1 because comparative correctness is partly robust to individual numeric noise.
The relative ordering across companies can still be right even when the exact values are not.

\subsection{Failure Analysis on BrowseComp+}
\label{app:bcp-failure-analysis}

BrowseComp+ questions are conjunctive ``needle'' queries: they identify a single person, place, work, or title through several oblique constraints that must all hold simultaneously.
For example, ``in a pre-2023 literary work by an author whose parent was the first professor of a field in their country, the protagonist is originally from [place]…'' (gold: Rebecca). 
The web-search agent scores low here (0.39 autonomous / 0.38 search-only). 
Its 63 failures in the search-only configuration split into two roughly equal patterns:

\begin{smartlist}
    \item \emph{Retrieval miss (non-answer): $33/63$.} Even when forced to search, the agent locates no source satisfying every constraint and explicitly declines.
    For example, it reports it ``could not find a single person who satisfies all parts of your query'' for the gold answers \emph{Kathryn Stack Morgan} and \emph{Laura Lojo-Rodríguez}. 
    Forcing an answer suppresses clarification requests but does not induce fabrication: the model still abstains honestly when retrieval fails, and the abstention is scored as incorrect.
    \item \emph{Wrong entity committed: $30/63$} the agent locks onto a specific candidate that satisfies a subset of the clues but is not the target, and cites a source for it (all 30 are grounded in at least one citation). 
    For example, it names the artist Amaarae (Ama Serwah Genfi) when the requested birth name was Emmanuel Kwesi Danso Arthur Junior.
\end{smartlist}

This contrasts with the LOONG failure mode (\autoref{app:loong-failure-analysis}): there the agent finds the correct entity but extracts the wrong figure, whereas on BrowseComp+ the entity itself is wrong or unfound. 
The difference stems from the task: BrowseComp+ requires satisfying many constraints jointly across the entire public web, so without the curated corpus that contains the annotated evidence document, the agent either surfaces no source meeting every constraint or commits to a partial match.

\section{Closed-Book Ablation and Correctness Decomposition}
\label{app:closed-book}

\autoref{fig:fine-grained-funnel} shows that answer correctness exceeds recall on LOONG L2 and KILT-HotpotQA, but not on BrowseComp+. 
To determine whether these gaps reflect the synthesizer compensating with parametric knowledge, or simply reflect questions that do not require complete retrieval to answer, we run a closed-book ablation: the same synthesizer model and prompt template as the main pipeline, but with the retrieved context entirely removed.

\paragraph{Setup.} We evaluate two closed-book conditions per dataset, both using the same synthesizer model (\texttt{gpt-5-mini}) and the same structured output schema as the main pipeline, but differing in what the synthesizer is told and given.

\para{no-guess} is the pipeline's default synthesizer prompt, run with the context field emptied:

\begin{quote}
\small
\textit{System:} You are a helpful assistant that answers questions based on
the provided context.\\
\textit{User:} Instructions:\\
1. Based on the following context answer the question.\\
2. Do not use any other information or your internal knowledge to answer the
question.\\
3. If required information is not present in the context, reply
correspondingly with ``Not enough context''\\[2pt]
Query: \{query\}\\
Context: \textit{(empty)}
\end{quote}

Because this prompt explicitly instructs the model \emph{not} to use its own knowledge, it measures the \emph{leak floor}: how often the model answers by disregarding that instruction, given nothing else to go on.

\para{best-guess} uses a separate
prompt that instructs the model to answer from its own knowledge instead of
declining:

\begin{quote}
\small
\textit{System:} You are a helpful assistant. Answer the question using your
own knowledge. Be concise and give the specific answer.\\
\textit{User:} \{question\}\\[2pt]
If you do not know the answer, reply exactly: I don't know.
\end{quote}

This measures the \emph{parametric ceiling}: the upper bound of what the model could answer from memory alone when explicitly asked to. 
Both conditions are scored with the same correctness metric used throughout the paper. 

The gap between them is informative in its own right: a low no-guess score paired with
a high best-guess score indicates the model holds latent knowledge of the answer.

\paragraph{Results.} ~\autoref{tab:closed-book} reports no-guess and best-guess
correctness alongside full-pipeline correctness, for all four benchmarks.

\begin{table}[h]
\centering
\begin{tabular}{lccc}
\toprule
Dataset & no-guess & best-guess & Pipeline correct \\
\midrule
LOONG L1     & 0.00 & 0.00 & 0.79 \\
LOONG L2     & 0.00 & 0.06 & 0.63 \\
KILT-HotpotQA & 0.00 & 0.67 & 0.86 \\
BrowseComp+  & 0.00 & 0.00 & 0.79 \\
\bottomrule
\end{tabular}
\caption{Closed-book correctness (no retrieved context) vs.\ full-pipeline
correctness. 
No-guess is scored under an instruction to decline when unsure;
Best-guess under an instruction to always guess.}
\label{tab:closed-book}
\end{table}

No-guess correctness is 0.00 on every benchmark: without being pushed to guess, the synthesizer never volunteers a confident, correct answer from memory alone under this instruction. 
This alone would suggest ``no parametric leakage under no context,'' but it understates the model's latent knowledge, since a model can hold information it is unwilling to assert outright. 

Best-guess correctness reveals the actual spread: near zero on LOONG (both levels) and BrowseComp+, but 0.67 on KILT.
This splits the four benchmark--level pairs into two groups: \emph{memory-free} (LOONG, BrowseComp+), where the model has no usable latent knowledge of the answer, and \emph{parametric-dominated} (KILT), where most of the pipeline's correctness is already recoverable without any retrieval at all.

\paragraph{LOONG L2: partial sufficiency, not memory.} With best-guess
correctness near zero, the 17 of 50 L2 queries that are answered correctly despite incomplete recall (\autoref{sec:funnel}) cannot be explained by memory. 
Manual inspection of these 17 shows two distinct patterns:
\begin{smartlist}
    \item \textbf{12 of 17}: the winning company's document was at least partially retrieved  (its name surfaced in context) but mostly not its value, let alone the full candidate set. 
    The synthesizer landed the comparison from the winner's name alone, without the deciding evidence itself.
    \item \textbf{5 of 17}: the winning company was absent from the retrieved text entirely — answered correctly purely from a rankable/robust partial subset (or luck). 
    For example, given a question of the form ``among candidate companies \{$A_1, \dots, A_n$\}, which had the highest [metric]?'' with gold winner \textsc{CirTran Corp}, 0
    of the 2 companies actually needed to resolve the comparison were retrieved, yet the answer was still correct. 
    Because the answer space is constrained to the small set of candidates named in the question itself, a correct answer here does not imply access to the correct underlying values.
    Instead, it can reflect the model recovering the right choice from an incomplete or even empty comparison set, which we term \emph{order-robust guessing}.
\end{smartlist}
Because best-guess closed-book correctness on LOONG L2 is near zero, this pattern cannot be attributed to parametric memory of the underlying financial figures: the model provably does not know these values unprompted.

\paragraph{KILT: shortcut and memory, entangled.} KILT's high best-guess
ceiling (0.67) reflects that HotpotQA questions, being sourced from Wikipedia, are frequently answerable from general knowledge.
Of the 34 of 100 queries answered correctly despite incomplete recall, we distinguish two patterns by whether the answer text is present anywhere in the retrieved
context:
\begin{itemize}
    \item \textbf{26 of 34}: the answer text is present in the retrieved context, even though not every annotated bridge hop was fetched. 
    This matches a well-documented artifact of the HotpotQA dataset, where many nominally multi-hop questions can be solved via a single-hop shortcut.
    For example, match salient entities in the question directly against a single retrieved passage without connecting the intermediate bridge~\cite{min2019compositional, jiang2019avoiding}. 
    For example, the question ``What film stars Rose McGowan and features the character of Earl
    McGraw's daughter?'' (gold answer: \emph{Planet Terror}) was answered correctly with only 1 of 3 annotated hops retrieved, because the answer text itself was present in what was retrieved.
    \item \textbf{8 of 34}: the answer text is absent from \emph{all} retrieved content, yet the query is answered correctly.
    For example: ``Who narrated the 26-episode series featuring the British national museum \ldots?'' (gold answer: \emph{Michael Redgrave}) and ``'Hold On' features a British singer whose debut album won what prize?'' (gold answer: \emph{Mercury Prize}), both answered correctly with 0 of 4 annotated hops retrieved and no occurrence of the answer text in retrieved content. 
    Here the final hop is never fetched at all, and the answer is supplied from parametric memory.
\end{itemize}

\paragraph{Summary.} ~\autoref{tab:summary-mechanism} restates the three
mechanisms identified in \autoref{sec:funnel} against the ablation
evidence supporting each.

\begin{table}[h]
\centering
\begin{tabular}{lll}
\toprule
Dataset & Mechanism & Ablation evidence \\
\midrule
BrowseComp+ & Retrieval-bound & Best-guess $=0.00$; recall $\approx$ correctness \\
LOONG L2 & Candidate-constrained guessing & Best-guess $\approx 0$; 5/17 correct with 0 relevant values retrieved \\
KILT & Shortcut + partial memory & Best-guess $=0.67$; 26/34 shortcut, 8/34 memory-only \\
\bottomrule
\end{tabular}
\caption{Correctness--recall divergence mechanism per dataset, with supporting evidence.}
\label{tab:summary-mechanism}
\end{table}

\section{Prompts}
\label{app:prompts}

For reproducibility, we list the exact prompts used at each pipeline stage.
All prompts use \texttt{gpt-5-mini} unless otherwise noted, and all synthesizer-facing prompts (synthesis, closed-book conditions) use the same structured output schema (\texttt{answer}, \texttt{relevant\_context\_present}).

\paragraph{Query processing.} Query decomposition and rephrasing uses a fixed template with structured output into a \texttt{TransformedQuery} object (\texttt{queries}, \texttt{query\_rephrased}, \texttt{query\_decomposed}).
In our benchmark runs this is computed once per query and cached, so the same decomposition is reused across all values of $k$ in the sweep in \autoref{fig:tradeoff-k-sweep}.

\begin{quote}
\small
\textit{System:} You are a helpful assistant that transforms user queries for websearch using the provided instructions\\
\textit{User:} Instructions:\\
1. Rephrase the query, IF AND ONLY IF, the current phrasing would not be good for search on a web search engine\\
2. IF it is a complex query, decompose it into a list of simpler web search queries that can provide the data necessary to answer the query\\
3. If the query was rephrased at all from the original query, set
\texttt{query\_rephrased} to true\\
4. If the query was decomposed, set \texttt{query\_decomposed} to true\\[2pt]
USER QUERY:\\
\{query\}
\end{quote}

\paragraph{Synthesis (main pipeline).} The synthesizer's answer-generation prompt (\Circled{Q4} in \autoref{fig:aistral-architecture}) is identical to the no-guess closed-book prompt (~\autoref{app:closed-book}), with one difference:
the context field is populated with the retrieved chunks rather than left empty.

\paragraph{Closed-book: no-guess.} See ~\autoref{app:closed-book}.

\paragraph{Closed-book: best-guess.} See ~\autoref{app:closed-book}.

\paragraph{Web-search agent baseline}
Both configurations use \texttt{gpt-5-mini} with the OpenAI Responses \texttt{web\_search} tool; they differ only in whether the tool is forced and whether the no-memory instruction is prepended to the question.

\begin{quote}
\small
\textbf{Autonomous:} tool\_choice left at default (the model decides whether to search); the question is passed as-is, with no instruction prepended.\\[4pt]
User: \texttt{<question>}
\end{quote}

\begin{quote}
\small
\textbf{Search-only (no memory):} tool\_choice forces the web\_search tool on every call; the following instruction is prepended to the question.\\[4pt]
User: You are a web-search answering agent. You MUST use the web\_search tool to look up the answer, and you must answer using ONLY information found in the retrieved web results. Do NOT rely on prior or parametric knowledge. Do NOT ask for clarification: if the question is ambiguous, choose the most reasonable interpretation and commit to a single, specific final answer. Be concise.\\
Question: \texttt{<question>}
\end{quote}

In both configurations, the agent's final message is taken directly as the answer, with no separate synthesis step.

\paragraph{Contextual chunking (+ctx) document summary.} At ingest, the \texttt{+ctx} variant generates one summary per document, cached per document and prepended verbatim to every chunk of that document.

\begin{quote}
\small
\textit{System:} You produce short retrieval-oriented context snippets.
Answer only with the requested context or summary.\\
\textit{User:} <document>\\
\{document\}\\
</document>\\[2pt]
Please give a short succinct summary of this document that can be added to each chunk of this document for the purposes of improving search retrieval of the chunks of this document.\\
Answer only with the summary and nothing else.
\end{quote}

\paragraph{Evaluation: closed-ended benchmarks.} We score LOONG, KILT-HotpotQA,
and BrowseComp+ with a single \texttt{GEval} metric,
\textsc{GoldCorrectness}, over (\texttt{INPUT}, \texttt{ACTUAL\_OUTPUT},
\texttt{EXPECTED\_OUTPUT}):

\begin{quote}
\small
Assess whether ACTUAL\_OUTPUT is correct relative to EXPECTED\_OUTPUT for
INPUT. Use EXPECTED\_OUTPUT as the authoritative reference answer and the
sole basis for correctness judgment. Compare the meaning of ACTUAL\_OUTPUT
against EXPECTED\_OUTPUT. Do not use outside knowledge, real-world
plausibility, or independent fact-checking to override or dispute
EXPECTED\_OUTPUT. Allow harmless differences in formatting, punctuation,
capitalization, commas, spacing, and minor unit wording when the underlying
answer meaning is unchanged. Accept missing currency symbols when the value
and meaning still clearly match EXPECTED\_OUTPUT, but treat a different or
implied different currency as incorrect. Treat currency meaning and numeric
sign as semantically important. Accept equivalent negative-number notation
such as a leading minus sign, accounting-style parentheses, or clearly
equivalent wording like a loss. Accept extra explanation if the core answer
remains clearly correct and is not contradicted.
\end{quote}

This single metric is a semantic match against the gold answer, lenient on formatting and currency notation but strict on numeric sign and currency identity.

\paragraph{Evaluation: open-ended benchmark (~\autoref{app:open-end-eval}).} 
The open-ended benchmark has no single gold answer, so we score each response along three separate \texttt{GEval} metrics over (\texttt{INPUT}, \texttt{ACTUAL\_OUTPUT}) only --- no expected output:

\begin{quote}
\small
\textbf{Correctness (GoldCorrectness):} Assess whether ACTUAL\_OUTPUT is factually correct and relevant to INPUT. [Judge] based on real-world plausibility and internal consistency. Penalize factual errors, fabricated claims, and contradictions. Do not require matching a [reference answer].\\[4pt]
\textbf{Completeness (QueryCompleteness):} Assess how completely ACTUAL\_OUTPUT answers the INPUT query by itself. [Reward covering all] essential requested points and penalize omissions of key parts of the question. 
Ignore stylistic differences.\\[4pt]
\textbf{Focus (Focus):} Assess how focused ACTUAL\_OUTPUT is on INPUT.
Higher score means the response is [concise], precise, and relevant to the query. Penalize digressions, unnecessary tangents, and irrelevant details.
\end{quote}

The dashed reference line in ~\autoref{fig:answer-quality-k-sweep} is \texttt{Opus}'s own gold answers, scored as \texttt{ACTUAL\_OUTPUT} against these same three prompts.

\end{document}